\documentclass[prl,preprintnumbers,amsmath,amssymb,superscriptaddress]{revtex4}

\usepackage{epsfig}
\usepackage{graphicx}
\usepackage{dcolumn}
\usepackage{bm}

\newcommand{\vk} {\vec{\bf k}}

\newcommand{\vf}{v_{\rm F}}

\newcommand{\epk}{\epsilon_{\bf \vec{k}}}

\newcommand{\HR}{-i \Delta_{\rm R}}
\newcommand{\HRm}{+i \Delta_{\rm R}}
\newcommand{\Hi}{\Delta_{\rm int}}

\begin{document}

\title{Spin-orbit coupling in curved graphene, fullerenes, nanotubes,
  and nanotube caps.}
\author{Daniel Huertas-Hernando}
\affiliation{Department of Physics, Norwegian University of Science and Technology, N-7491, Trondheim, Norway}
\author{F. Guinea}
\affiliation{Instituto de Ciencia de Materiales de Madrid, CSIC, Cantoblanco E28049 Madrid, Spain}
\author{Arne Brataas}
\affiliation{Department of Physics, Norwegian University of Science and Technology, N-7491, Trondheim, Norway}
\affiliation{Centre for Advanced Study, Drammensveien 78, 0271 Oslo, Norway}

\begin{abstract}
A continuum model for the effective spin orbit
interaction in graphene is derived from a tight-binding model which includes
the $\pi$ and $\sigma$ bands. We analyze the combined effects of the
intraatomic spin orbit coupling, curvature, and applied electric field,
using perturbation theory. We recover the effective spin-orbit Hamiltonian derived 
recently from group theoretical arguments by Kane and Mele. We find, for flat graphene, that 
the intrinsic spin-orbit coupling $\Hi \propto \Delta^ 2$ and the Rashba
coupling due to an perpendicular electric field $\cal E$, $\Delta_{\cal E}
\propto \Delta$, where $\Delta$ is the intraatomic spin-orbit coupling
constant for carbon.  Moreover we show that local curvature of the graphene  
sheet induces an extra spin-orbit coupling term $\Delta_{\rm curv} \propto
\Delta$. For the values of $\cal E$  and curvature profile reported in actual 
samples of graphene, we find that $\Hi  < \Delta_{\cal E} \lesssim
\Delta_{\rm curv}$. The effect of spin orbit coupling on derived materials of 
graphene like fullerenes, nanotubes, and nanotube caps, is also studied. For fullerenes, only $\Hi$ is
important. Both for nanotubes  and nanotube caps $\Delta_{\rm curv}$ is in the order of 
a few Kelvins. We reproduce the known appearance of a gap and
spin-splitting in the energy spectrum of nanotubes due to the 
spin-orbit coupling. For nanotube caps, spin-orbit coupling causes
spin-splitting of the localized states at the cap, which could 
allow spin-dependent field-effect emission.   
\end{abstract}


\maketitle
\section{Introduction.}
  A single layer of carbon atoms in a honeycomb lattice, \emph{graphene},
is an interesting two-dimensional system due to its remarkable low energy electronic 
properties\cite{W47,SW58,mele}, \emph{e.g.} a zero density of states 
at the Fermi level without an energy gap, and a linear, rather than
parabolic, energy dispersion around the Fermi level. The electronic
properties of the many realizations of the honeycomb lattice of carbon 
such, \emph{e.g.} bulk graphite (3D), carbon nanotube wires (1D), 
carbon nanotube quantum dots (0D), and curved surfaces such as fullerenes, 
have been studied extensively during the last decade. 
However, its two dimensional (2D) version, graphene, a stable atomic layer of carbon atoms, 
remained for long ellusive among the known crystalline structures of carbon. 
Only recently, the experimental realization of stable, 
highly crystalline, single layer samples of
graphene \cite{Netal04,Netal05,Netal05b,Zetal05b}, 
have been possible. Such experimental developments have generated a renewed interest 
in the field of two dimensional mesoscopic systems. The peculiar
electronic properties of graphene are quite different from that of 2D 
semiconducting heterostructures samples. It has been
found that the integer Hall effect in graphene is different than the
``usual'' Quantum Hall effect in semiconducting structures\cite{NGP05,PGN06,GS05,BF06}. 
Moreover, it has been theoretically suggested that a variety of properties, 
\emph{e.g.} weak (anti)localisation\cite{K05,MG06,Metal06,MKFSAA06,AAEF06,ALTH06},
 shot-noise\cite{TTTRB06} and anomalous 
tunneling-Klein's paradox\cite{KNG06}, are qualitatively different from the 
behavior found in other 2D systems during the last decades. All these 
predictions can now be directly investigated by experiments. The activity  
in graphene, both theoretically and experimentally, is at present very
intense. However so far, the work has mainly focused on i) the fact 
that the unit cell is described by two inequivalent triangular 
sublattices $A$ and $B$ intercalated, and  ii) there are two independent $k$-points, 
$K$ and $K'$, corresponding to the two inequivalent corners of the Brillouin zone
of graphene.  
The Fermi level is located at these K and K'
points and crosses the $\pi$ bands of graphene (see  Fig.[\ref{sigma_Bnds}] for details). 
These two features provide an exotic fourfold degeneracy of the low
energy (spin-degenerate) states of graphene. These states can be described 
by two sets of two-dimensional chiral spinors which follow the massless
Dirac-Weyl equation and describe the electronic states of the system near the K and K' 
points where the Fermi level is located. 
Neutral graphene has one electron per carbon atom in the $\pi$ band, 
so the band below the Fermi level is full (electron-like states) 
and the band above it is empty (hole-like states). 
Electrons and holes in graphene behave like relativistic Dirac fermions. 
The Fermi level can be moved by a gate voltage underneath the graphene
sample\cite{Netal04}. State-of-the-art samples are very clean, with 
mobilities $\mu \sim 15 000 cm^2 V^-1 s^-1$\cite{Netal05b}, so 
charge transport can be ballistic for long distances across the sample. 
From the mobilities of the actual samples, it is believed that impurity scattering is weak. 
Furthermore, it has been recently suggested that the
chiral nature of graphene carriers makes disordered regions
transparent for these carriers independently of the disorder, as long as it
is smooth on the scale of the lattice constant\cite{SA02,MG06,Metal06}. 

Less attention has been given to the spin so far. The main interactions that could affect the spin degree of freedom 
in graphene seem to be the spin-orbit coupling and exchange interaction. 
It is not known to which extent magnetic impurities are present in actual graphene samples. Their effect seem small though, as noticed recently when investigating weak localization and universal conductance fluctuations in graphene\cite{Metal06}.
Spin-orbit interaction in graphene is supposed to be
weak, due to the low atomic number $Z=6$ of carbon. 
Therefore both spin splitting and spin-flip due to the
combination of spin-orbit and scattering due to disorder is supposed to be
not very important. As a result, the spin degree of freedom is assumed to
have a minor importance and spin degenerate states are assumed. 
Besides, the spin degeneracy is considered to be ``trivial'' in comparison 
to the fourfold degeneracy previously mentioned, described by a pseudo-spin 
degree of freedom. At present, there is a large activity in the study of the dynamics of this 
pseudo-spin degree of freedom\cite{NGP05,PGN06,GS05,BF06,K05,MG06,Metal06,MKFSAA06,KNG06,SA02,KNMcD06,AAEF06,ALTH06,TTTRB06,Katsn06,KhvE06,KM05,SSTH05,SHMSM06}. 

We think that the physics of the electronic spin in
graphene must be investigated in some detail, however. Although it could be that the electronic spin is 
not as important/exotic as the pseudo-spin when studying bulk properties, edge states may be quite different. Induced magnetism at the edges of the surface of graphite samples irradiated with protons have been reported\cite{esquinazi}. Moreover, perspectives of spintronic 
applications in graphene could be very promising, so it is important to
clarify the role of the electronic spin. This is one of the main
purposes of the present paper. Moreover, we feel that the existent knowledge
about the spin-orbit interaction in graphene is not yet complete\cite{KM05} and that certain, 
both quantitative and qualitative, points must be discussed in more detail. 
That is why we focus our discussion on the spin-orbit coupling. The effect of 
other interactions as the exchange interaction will be discussed elsewhere.

Spin-orbit coupling in graphene has an intrinsic
part, completely determined from the symmetry properties of the 
honeycomb lattice. This is similar to the Dresselhauss spin-orbit interaction in
semiconducting heterostructures\cite{DSS55}. Group theoretical arguments allow to obtain 
the form of the effective Hamiltonian for the intrinsic spin-orbit coupling
around the $K$, $K'$ points\cite{KM05,S55, DD65}. 
It was predicted that this interaction opens-up  a gap $\Hi$ in the energy 
dispersion. However, the strength of this intrinsic spin-orbit coupling is 
still a subject of discussion, although it is believed to be rather small, 
due to the weakness of the atomic intraatomic spin orbit coupling of
carbon $\Delta$.  If an electric field $\cal E$ is applied perpendicular to the
sample, a Rashba interaction\cite{BYCHRASH84} $\Delta_{\cal E}$ will also be present in graphene. 
Analogous to the intrinsic coupling, group theoretical arguments allow 
deducing the form of the Rashba interaction\cite{KM05,SSTH05}. The strength of this Rashba spin-orbit coupling is also still under discussion. 

We follow a different approach. We set up a tight binding model where 
we consider both the $\pi$ and $\sigma$ bands of graphene and the intraatomic spin-orbit coupling $\Delta$. We also include 
curvature effects of the graphene surface and the presence of a perpendicular electric field $\cal E$. 
Starting from this model, we obtain an effective Hamiltonian for the $\pi$ bands, by second order perturbation theory, which is formally the same as the effective Hamiltonian obtained previously from group theoretical methods\cite{S55, DD65} by Kane and Mele\cite{KM05}. Moreover, we show that curvature effects between nearest neighbor atoms introduce 
an extra term $\Delta_{\rm curv}$ into the effective spin-orbit interaction of graphene, similar to the Rashba interaction due to the electric field $\Delta_{\cal E}$. We obtain explicit expressions for these three couplings in terms of band structure parameters. Analytical expressions and numerical estimates are given in Table[\ref{numerical}]. 

We find that the intrinsic interaction $\Hi \sim 10$mK is two orders of magnitude smaller than what was recently estimated\cite{KM05}. Similar estimates for $\Hi$  have been reported recently\cite{SHMSM06,YAOYEFANG06,HMING06}. Moreover, we find that for typical values of the electric field as \emph{e.g.} used by Kane and Mele\cite{KM05} $\Delta_{\cal E} \sim 70$mK. Similar discussion for $\Delta_{\cal E}$ has appeared also recently\cite{HMING06}. So spin-orbit coupling for flat graphene is rather weak.  Graphene samples seem to have an undulating surface however\cite{Metal06}. Our estimate for the typical observed ripples indicates that $\Delta_{\rm curv} \sim 0.2$K. It seems that curvature effects on the scale of the distance between neighbouring atoms could increase the strength of the spin-orbit coupling at least one order of magnitude with respect that obtained for a flat surface. More importantly,  this type of  ``intrinsic'' coupling will be present in graphene as long as its surface is corrugated even if  $\cal E=$0 when $\Delta_{\cal E}=$0.

 The paper is organized as follows: The next section presents a tight binding hamiltonian for the band
structure and the intraatomic spin-orbit coupling, curvature
effects and a perpendicular electric field. Then, the three
  effective spin-orbit couplings $\Hi, \Delta_{\cal E},\Delta_{\rm curv}$ for a continuum model 
of the spin-orbit interaction for the $\pi$ bands in graphene at the $K$ and $K'$ points are derived. 
Estimates of the values are given at the end of the
section. The next section applies the effective spin-orbit hamiltonian to:
i) fullerenes, where it is shown that spin-orbit coupling effects play a small role at 
low energies ii) nanotubes, where known results are recovered and iii)
nanotubes capped by semispherical fullerenes, where it is shown that the
spin orbit coupling can lead to localized states at the edges of the bulk
 subbands. The last calculation includes also a continuum model for the
electronic structure of nanotube caps, which, to our knowledge, has not
been discussed previously. A section with the main conclusions completes
the paper.

\section{Derivation of continuum models from intraatomic interactions.}
\subsection{Electronic bands.}

\begin{figure}
\begin{center}
\includegraphics*[width=12cm,angle=-90]{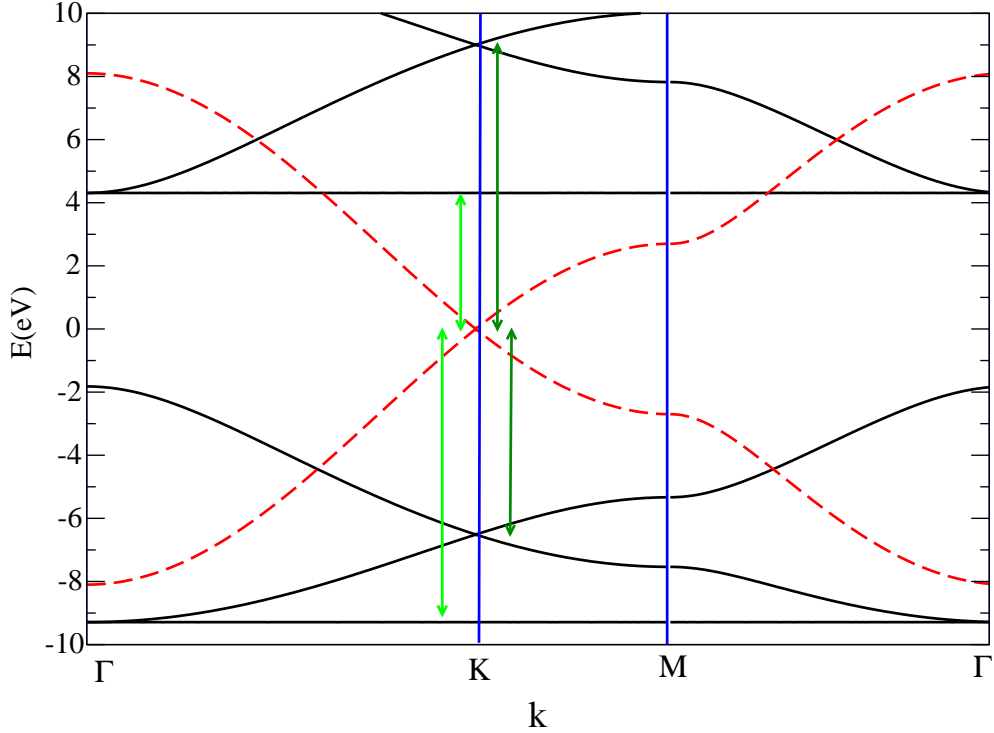}
\caption{``(Color online)'': Black (Full) curves: $\sigma$ bands. Red (Dashed) curves: $\pi$ bands. 
The dark and light green (grey) arrows give contributions to the up
and down spins at the A sublattice respectively. The opposite contributions
  can be defined for the B sublattice. These interband transitions are equivalent to the processes depicted in
  Fig.[\protect{\ref{intrinsic_hoppings}}].}
\label{sigma_Bnds}
\end{center}
\end{figure}

The orbitals corresponding to the $\sigma$ bands of graphene are made by
linear combinations of the $2s$, $2p_x$ and $2p_y$ atomic orbitals, whereas
the orbitals of the $\pi$ bands are just the $p_z$ orbitals. 
We consider the following Hamiltonian:
\begin{equation} 
{\cal H} = {\cal H}_{\rm SO} + 
{\cal H}_{\rm atom }  + {\cal H}_{\pi} + {\cal H}_{\sigma}
\label{hamil} \, ,
\end{equation}
where the atomic hamiltonian in the absence of spin orbit coupling is:
\begin{equation}
{\cal H}_{\rm atom } =\epsilon_p \sum_{i=x,y,z; s'=\uparrow , \downarrow } c^\dag_{i  s'} 
c_{i  s'}+\epsilon_{s} \sum_{ s ; s'=\uparrow , \downarrow} c^\dag_{s , s'}
c_{s , s'}.
\label{hamil_ion}
\end{equation}
where $\epsilon_{p,s}$ denote the atomic energy for the $2p$ and $2s$ atomic orbitals of carbon, the operators $c_{i;s'}$ and $c_{s;s'}$ refer to $p_z$, $p_x$, $p_y$ and $s$ atomic orbitals respectively and $s'= \uparrow, \downarrow$ denote the electronic spin. 
${\cal H}_{\rm SO}$ refers to the atomic spin-orbit coupling occuring at the carbon atoms and the terms ${\cal H}_{\pi},{\cal H}_{\sigma} $ 
describe the $\pi$ and $\sigma$ bands. In the following, we will set our origin of energies such that $\epsilon_{\pi}=0$. We use a nearest neighbor
hopping model between the $p_z$ orbitals for ${\cal H}_{\pi}$, using one parameter $V_{pp
  \pi}$. The rest of the intraatomic hoppings are the nearest
neighbor interactions $V_{pp \sigma} , V_{sp \sigma}$ and $V_{ss \sigma}$ between the atomic orbitals $s,p_x,p_y$ of the $\sigma$ band. 
We describe the $\sigma$ bands using a variation of an
analytical model used for three dimensional semiconductors with the diamond
structure\cite{TW71}, and which was generalized to the related problem of the
calculation of the acoustical modes of graphene\cite{G81}. The model for the
sigma bands is described in Appendix A. The bands can be calculated
analytically as function of the parameters:
\begin{eqnarray}
V_1 &= &\frac{\epsilon_s - \epsilon_p}{3} \nonumber \\
V_2 &= &\frac{2 V_{pp \sigma} + 2 \sqrt{2} V_{sp \sigma} + V_{ss \sigma}}{3}.
\end{eqnarray} 
The band structure for graphene is shown in Fig.(\ref{sigma_Bnds}).

\subsection{Intraatomic spin-orbit coupling.}
The intraatomic spin orbit coupling is given by ${\cal H}_{\rm SO} = \Delta
  \vec{\bf L} \vec{\bf s} ~$ \cite{BJ83} where $\vec{\bf L}$ and $\vec{\bf s}$ are, the total atomic angular momentum operator and total electronic spin operator respectively, and $\Delta$ is the intraatomic spin-orbit coupling constant. 
We define:
\begin{eqnarray}
s_+ &\equiv &\left( \begin{array}{cc} 0 &1 \\ 0 &0 \end{array} \right) 
\nonumber\\ s_- &\equiv &\left( \begin{array}{cc} 0 &0 \\ 1 &0 \end{array} \right)
\nonumber\\ s_z &\equiv &\left( \begin{array}{cc} \frac{1}{2} &0 \\ 0
    &-\frac{1}{2} \end{array} \right) \nonumber \\ L_+ &\equiv &\left(
  \begin{array}{ccc} 0 &\sqrt{2} &0 \\ 0 &0 &\sqrt{2} \\ 0 &0 &0 \end{array}
\right) \nonumber \\ L_- &\equiv &\left(
  \begin{array}{ccc} 0 &0 &0 \\ \sqrt{2} &0 &0 \\ 0 &\sqrt{2} &0 \end{array}
\right) \nonumber \\ L_z &\equiv &\left( \begin{array}{ccc} 1 &0 &0 \\ 0 &0
    &0 \\ 0 &0 &-1 \end{array} \right) \nonumber \\ | p_z \rangle &\equiv &| L
    = 1 , L_z = 0 \rangle \nonumber \\ | p_x \rangle &\equiv
    &\frac{1}{\sqrt{2}} \left( | L=1 , L_z = 1 \rangle + | L=1 ,L_z
    = -1 \rangle \right) \nonumber \\ | p_y \rangle &\equiv
    &\frac{+ i}{\sqrt{2}} \left( | L=1 , L_z = 1 \rangle - | L=1 ,L_z
    = -1 \rangle \right),
\label{definitions}
\end{eqnarray}
 Using these definitions, the intraatomic spin-orbit hamiltonian becomes:
\begin{equation}
{\cal H}_{\rm SO} = \Delta \left[ \frac{L_+ s_- + L_- s_+}{2} + L_z s_z \right]
\label{SO}
\end{equation}
The Hamiltonian Eq. (\ref{SO}) can be written in second quantization language as:
\begin{equation}
{\cal H}_{\rm SO} = \Delta \left[ c^\dag_{z \uparrow} c_{x \downarrow} -
  c^\dag_{z \downarrow} c_{x \uparrow} + i c^\dag_{z \uparrow} c_{y
  \downarrow} - i c^\dag_{z \downarrow} c_{y \uparrow} + i c^\dag_{x
  \downarrow} c_{y \downarrow} - i c^\dag_{x\uparrow} c_{y \uparrow}+ h. c. \right].
\end{equation}
where the operators $c^\dag_{z,x,y;s'}$ and $c_{z,x,y;s'}$ refer to the corresponding $p_z$, $p_x$ and $p_y$ atomic orbitals. The intraatomic hamiltonian is a $6 \times 6$ matrix which can be split into
two $3 \times 3$ submatrices:
\begin{equation}
{\cal H}_{\rm SO} =\left( \begin{array}{cc} {\cal H}^{11}_{\rm SO} &0 \\ 0
    & {\cal H}^{22}_{\rm SO} \end{array} \right)
\end{equation}
The block ${\cal H}^{11}_{\rm SO}$ acts on the basis states $| p_z \uparrow \rangle , | p_x \downarrow \rangle$ and $| p_y \downarrow \rangle$:
\begin{equation}
{\cal H}^{11}_{\rm SO} = \frac{\Delta}{2} \left( \begin{array}{ccc} 0 &1 &  i \\ 1
    &0 & -i \\ - i
    & i &0 \end{array} \right).
\end{equation}
On the other hand ${\cal H}^{22}_{\rm SO}$:
\begin{equation}
{\cal H}^{22}_{\rm SO} = \frac{\Delta}{2} \left( \begin{array}{ccc} 0 &- 1 & i \\ -1
    &0 &-i \\-  i
    &i &0 \end{array} \right)
\end{equation}
acts on $| p_z \downarrow \rangle , | p_x \uparrow \rangle$ and
$| p_y \uparrow \rangle$ states. The eigenvalues of these $3 \times 3$ matrices are
$+ \Delta$ $( J = 3/2 )$ which is singly degenerate and $- \Delta / 2 $ $( J = 1/2 )$ which is doubly degenerate.


The term $L_+ s_- + L_- s_+$ of the intraatomic spin orbit coupling Hamiltonian, Eq.(\ref{SO}), allows for transitions between states 
of the $\pi$ band near the $K$ and $K'$ points of the Brillouin zone, with states
from the $\sigma$ bands at the same points. These transitions imply a change of the electronic 
degree of freedom, \emph{i.e.} a ``spin-flip'' process. 

We describe the $\sigma$
bands by the analytical tight binding model presented in Appendix A. The six
$\sigma$ states at the $K$ and $K'$ points can be split into two Dirac
doublets, which disperse linearly, starting at energies 
$\epsilon_\sigma ( K , K' ) = V_1 / 2 \pm \sqrt{9 V_1^2 / 4 +
  V_2^2}$ and two flat bands at $\epsilon_\sigma ( K , K' ) = -V_1 \pm
V_2$. We denote the two $\sigma$ Dirac spinors as $\psi_{\sigma 1}$ and
$\psi_{\sigma 2}$, and the two other ``flat'' orbitals as $\phi_{\sigma 1}$ and
$\phi_{\sigma 2}$. The intraatomic spin orbit hamiltonian for the $K$ and $K'$ point 
becomes:
\begin{eqnarray}
{\cal H}_{SO K} &\equiv &\frac{\Delta}{2} \int d^2 \vec {\bf r} \sqrt{\frac{2}{3}}
    \left\{  \cos \left( \frac{\alpha}{2}  \right) \left[ \Psi^\dag_{A K
    \uparrow} ( \vec {\bf r} ) \psi_{\sigma 1 A K \downarrow} ( \vec {\bf r}
    ) + \Psi^\dag_{B K \uparrow} ( \vec {\bf r} ) \psi_{\sigma 1 B K \downarrow} ( \vec {\bf r} ) \right] +
    \right. \nonumber \\ 
&+ &\left. \sin \left( \frac{\alpha}{2} \right) \left[ \Psi^\dag_{A K
    \uparrow} ( \vec {\bf r} ) \psi_{\sigma 2 A K \downarrow} ( \vec {\bf r}
    ) + \Psi^\dag_{B K \uparrow} ( \vec {\bf r} ) \psi_{\sigma 2 B K \downarrow} ( \vec {\bf r} ) \right] \right\}
 + 
\sqrt{\frac{2}{3}} \left[ \Psi^\dag_{A K \uparrow}  ( \vec {\bf r} ) +
    \Psi^\dag_{B K \uparrow} (
  \vec {\bf r} ) \right] \phi_{1 \downarrow} ( \vec {\bf r} ) 
 + h. c. \nonumber \\
{\cal H}_{SO K'} &\equiv &\frac{\Delta}{2} \int d^2 \vec {\bf r} \sqrt{\frac{2}{3}}
    \left\{  \cos \left( \frac{\alpha}{2}  \right) \left[ \Psi^\dag_{A K'
    \uparrow} ( \vec {\bf r} ) \psi_{\sigma 2 A K' \downarrow} ( \vec {\bf r}
    ) + \Psi^\dag_{B K' \uparrow} ( \vec{\bf r} ) \psi_{\sigma 2 B K' \downarrow} ( \vec {\bf r} ) \right] +
    \right. \nonumber \\ 
&+ &\left. \sin \left( \frac{\alpha}{2} \right) \left[ \Psi^\dag_{A K'
    \uparrow} ( \vec {\bf r} ) \psi_{\sigma 1 A K' \downarrow} ( \vec {\bf r}
    ) + \Psi^\dag_{B K' \uparrow} ( \vec {\bf r} ) \psi_{\sigma 1 B K' \downarrow} ( \vec {\bf r} ) \right] \right\}
 + \sqrt{\frac{2}{3}} \left[ \Psi^\dag_{A K' \uparrow}  ( \vec {\bf r} ) +
    \Psi^\dag_{B K' \uparrow} (
  \vec {\bf r} ) \right] \phi_{2 \downarrow} ( \vec {\bf r} ) 
 + h. c.
\label{hamil_pi_sigma}
\end{eqnarray}
where $\Psi$ stands for the two component spinor of the $\pi$ band, and $\cos
( \alpha / 2 )$ and $\sin ( \alpha / 2 )$ are matrix elements given by:
\begin{equation}
\alpha = \arctan \left[ \frac{( 3 V_1 ) / 2}{\sqrt{( 9 V_1^2 ) / 4 + V_2^2} }
\right].
\end{equation} 

\begin{figure}
\begin{center}
\includegraphics*[width=10cm]{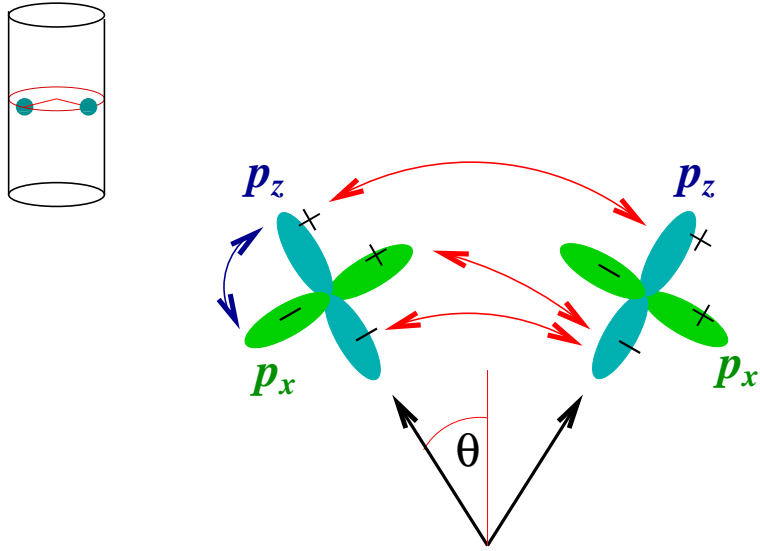}
\caption{``(Color online)'': Sketch of the relevant orbitals, $p_x$ and $p_z$ needed for the analysis of spin-orbit effects in a curved nanotube. The arrows stand for the different hoppings described in the text.}
\label{nanotube}
\end{center}
\end{figure}

Next we would like to consider two posibilities: 
i) A curved graphene surface. 
ii) The effect of a perpendicular electric field applied 
to flat graphene. In the latter case we will have to consider another intraatomic process besides the intraatomic spin-orbit coupling, the atomic Stark effect. 

\subsection{Effects of curvature.}
In a curved graphene sheet, a hopping between the orbitals in the $\pi$ and
$\sigma$ bands is induced\cite{A00}. First we will use a simple geometry 
to illustrate the effect of curvature between neighbouring atoms. This geometry is schematically shown in
Fig.[\ref{nanotube}], where we first consider two atoms  at the same height along the axis of the tube. 
In this geometry we consider that the $p_z$ orbitals are oriented normal to the surface of the nanotube, the $p_x$ orbitals are oriented along the surface circumference (Fig.[\ref{nanotube}] ) and the $p_y$ orbitals are 
parallel to the nanotube axes. The curvature modifies the hopping between 
the two atoms compared to the flat surface for the $p_z$ and $p_x$ orbitals 
but will not change, for this simple case, the hopping between $p_y$ orbitals.   The (reduced)
$p_z$-$p_x$ hopping hamiltonian is the sum of three terms:
\begin{eqnarray}
{\cal H}_{\rm T} &= &\sum_{s'} \left[ V_{pp \pi} \cos^2 ( \theta ) + V_{pp \sigma} \sin^2 (
\theta) \right] c^\dag_{z 1 s'} c_{z 0 s'} - \left[ V_{pp \pi} \sin^2 ( \theta ) +
V_{pp \sigma} \cos^2 ( \theta ) \right] c^\dag_{x 1 s'} c_{x 0 s'} + \nonumber \\ &+
&V_{sp \sigma} \sin^2 ( \theta ) c^\dag_{z 1 s'} c_{s 0 s'} + \sin ( \theta )
\cos ( \theta ) \left( V_{pp \pi} - V_{pp \sigma} \right) \left( c^\dag_{z 1 s'} c_{x 0 s'} -
    c^\dag_{x 1 s'} c_{z 0 s'} \right) + h. c.
\label{hopping_eff}
\end{eqnarray}
where $0$ and $1$ denote the two atoms considered and $\theta$ is the angle between the fixed $Z$ axis and the direction normal to the curved surface (Fig.[\ref{nanotube}] ). The angle $\theta$, in the limit when the radius of curvature is much longer
than the interatomic spacing, $a \ll R$, is given by $\theta \approx a /
R$. 

The hopping terms induced by (intrinsic) curvature discussed here break
the isotropy of the lattice and lead to an effective anisotropic coupling
between the $\pi$  and $\sigma$ bands in momentum space.

The previous discussion can be extended to the case of general curvature when the graphene sheet has two different curvature
radii, $R_1$ and $R_2$ corresponding to the $x$ and $y$ directions in the plane. In that case, the factor $R^{-1}$ has to be replaced by $R_1^{-1} + R_2^{-1}$. We now expand on $\theta \sim a/R_{1,2} \ll 1$. By projecting onto the Bloch wavefunctions of the
$\pi$ and $\sigma$ bands at the $K$ and $K'$ points, we find:
\begin{eqnarray}
{\cal H}_{T K} &\equiv & \left( V_{pp \sigma} - V_{pp \pi} \right)
\sqrt{\frac{3}{2}}
\left( \frac{a}{R_1} + \frac{a}{R_2} \right) \int d^2 \vec{\bf r} 
    \left\{  \cos \left( \frac{\alpha}{2}  \right) \left[ \Psi^\dag_{A K
    \uparrow} ( \vec{\bf r} ) \psi_{\sigma 1 B K \uparrow} ( \vec{\bf r}
    ) + \Psi^\dag_{B K \uparrow} ( \vec {\bf r} ) \psi_{\sigma 1 A K \uparrow} ( \vec{\bf r} ) \right] +
    \right. \nonumber \\ 
&+ &\left. \sin \left( \frac{\alpha}{2} \right) 
\left[ \Psi^\dag_{A K
   \uparrow} ( \vec{\bf r} ) \psi_{\sigma 2 B K \uparrow} ( \vec{\bf r}
    ) + \Psi^\dag_{B K \uparrow} ( \vec {\bf r} ) \psi_{\sigma 2 A K \uparrow} ( \vec{\bf r} ) 
\right] \right\}  + h. c.
\label{hamil_curv}
\end{eqnarray}
and a similar expression for the $K'$ point.

The induced spin orbit coupling, however, includes only contributions from the four $\sigma$
bands at $K$ and $K'$ with $\epsilon_\sigma = V_1/2 \pm \sqrt{(9 V_1^2)/4 +
  V_2^2}$, as those are the only bands coupled to the $\pi$ band by the
intraatomic spin orbit term considered here, Eq.(\ref{hamil_pi_sigma}).
We now assume that the energies of the $\sigma$ bands are well separated from
the energy of the $\pi$ bands ($\epsilon_\pi = 0$ at the $K$ and $K'$
points). Then, we can use second order perturbation theory and from
Eq.(\ref{hamil_pi_sigma}) and Eq.(\ref{hamil_curv}) we obtain an effective
hamiltonian acting on the states of the $\pi$ band.
\begin{equation}
{\cal H}_{{\rm curv} K \pi} \equiv -i \frac{\Delta ( V_{pp
  \sigma} - V_{pp \pi} ) V_1}{2 V_1^2 + V_2^2} \left( \frac{a}{R_1} + \frac{a}{R_2} \right)
  \int d^2 \vec{\bf r} \left(\Psi^\dag_{A K \uparrow} (
  \vec{\bf r} ) \Psi_{B K \downarrow} ( \vec{\bf r} ) -  \Psi^\dag_{ B K \downarrow} \Psi_{A K \uparrow} \right) . \nonumber
\end{equation}
\begin{equation}
{\cal H}_{{\rm curv} K' \pi} \equiv -i \frac{\Delta ( V_{pp
  \sigma} - V_{pp \pi} ) V_1}{2 V_1^2 + V_2^2} \left( \frac{a}{R_1} + \frac{a}{R_2} \right)
  \int d^2 \vec{\bf r} \left(-\Psi^\dag_{A K' \downarrow} (
  \vec{\bf r} ) \Psi_{B K' \uparrow} ( \vec{\bf r} ) +  \Psi^\dag_{ B K' \uparrow} \Psi_{A K' \downarrow} \right).
\label{RashbaCURV}
\end{equation}

\subsection{Effect of an electric field.}
Now we discuss the atomic Stark effect due to a perpendicular electric field
${\cal E}$.  In this case, we need to consider the $| s \rangle$ orbital of the $\sigma$ bands at each site, and
the associated hopping terms. The hamiltonian for this case includes the couplings:
\begin{equation}
{\cal H}_{\cal E} = \sum_{i=1,2;s'= \uparrow , \downarrow} \left( \lambda {e \cal
  E}  c^\dag_{is;s' } c_{iz;s'} + \epsilon_s c^\dag_{is;s'}
  c_{is;s'} + h. c. \right) + V_{sp \sigma}
  \sum_{s'= \uparrow , \downarrow} \left( 
  a_x c^\dag_{1 x ;s'} c_{0 s ;s'} + a_y c^\dag_{1 y ;s'} c_{0 s ;s'} +
  h. c. \right)
\end{equation}
where $\lambda = \langle p_z | \hat{z} | s \rangle$ is a electric dipole 
transition which induces hybridization between the $s$ and $p_z$ orbitals and where
$a_x$ and $a_y$ are the $x$ and $y$ components of the vector connecting
the carbon atoms 0 and 1. 
First, we consider the situation $a_x=1$ and $a_y=0$. Again $V_{sp \sigma}$ is the hopping integral between the $2s$ and $2p_x, 2p_y$ atomic orbitals corresponding to the $\sigma$ band. We can now have processes such as:
\begin{eqnarray}
| p_z 0 \uparrow \rangle &\xrightarrow{\cal E} | s 0 \uparrow
\rangle  &\xrightarrow{V_{sp \sigma}} | p_x 1 \uparrow \rangle
\xrightarrow{\Delta} | p_z 1 \downarrow \rangle \nonumber \\
| p_z 0 \uparrow \rangle  &\xrightarrow{\Delta} | p_x 0 \downarrow
\rangle &\xrightarrow{V_{sp \sigma}} | s 1 \downarrow \rangle \xrightarrow{\cal
  E} | p_z 1 \downarrow \rangle
\label{hopping_eff_3}
\end{eqnarray}
The intermediate orbitals $| s 0 \rangle$ and $| p_x 1 \rangle$ are part
of the sigma bands. As before, we describe them using the analytical fitting discussed
in Appendix A. The $| s 0 \rangle$ is part of the dispersive bands, and it
has zero overlap with the two non dispersive $\sigma$ bands.
The processes induced by the electric field, in momentum space, lead finally
to:
\begin{eqnarray}
{\cal H}_{{\cal E} K} &\equiv & 
\lambda {e \cal E} \sqrt{\frac{1}{3}} \int d^2 \vec{\bf r}
    \left\{  \sin \left( \frac{\alpha}{2}  \right) \left[ \Psi^\dag_{A K
    \uparrow} ( \vec{\bf r} ) \psi_{\sigma 1 A K \uparrow} ( \vec{\bf r}
    ) + \Psi^\dag_{B K \uparrow} ( {\bf
    \vec{r}} ) \psi_{\sigma 1 B K \uparrow} ( \vec{\bf r} ) \right] +
    \right. \nonumber \\ 
&+ &\left. \cos \left( \frac{\alpha}{2} \right) 
\left[ \Psi^\dag_{A K
   \uparrow} ( \vec{\bf r} ) \psi_{\sigma 2 A K \uparrow} ( \vec{\bf r}
    ) + \Psi^\dag_{B K \uparrow} ( {\bf
   \vec{r}} ) \psi_{\sigma 2 B K \uparrow} ( \vec{\bf r} ) 
\right] \right\}  + h. c.
\label{hamil_field}
\end{eqnarray}
 Note that this hamiltonian
mixes the states in the $\pi$ band with states in the $\sigma$ bands which
are orthogonal to those in Eq.(\ref{hamil_pi_sigma}). Combining Eq.(\ref{hamil_field}) and
Eq.(\ref{hamil_pi_sigma}) we obtain, again by second order perturbation theory, an effective hamiltonian for the $\pi$
band:
\begin{equation}
{\cal H}_{{\cal E} K \pi}  \equiv - i \frac{2 \sqrt{2}}{3} \frac{ \Delta \lambda
  {e \cal E} V_2}{2 V_1^2 + V_2^2} \int d^2 \vec{\bf r} \left(\Psi^\dag_{A K \uparrow} (
  \vec{\bf r} ) \Psi_{B K \downarrow} ( \vec{\bf r} ) -  \Psi^\dag_{ B K \downarrow} \Psi_{A K \uparrow} \right).\nonumber
\end{equation}
\begin{equation}
{\cal H}_{{\cal E} K' \pi}  \equiv -i \frac{2 \sqrt{2}}{3} \frac{ \Delta \lambda
  {e \cal E} V_2}{2 V_1^2 + V_2^2}  \int d^2 ( \vec{\bf r} ) \left(-\Psi^\dag_{A K' \downarrow} (
  \vec{\bf r} ) \Psi_{B K' \uparrow} ( \vec{\bf r} ) +  \Psi^\dag_{ B K' \uparrow} \Psi_{A K' \downarrow} \right).
\label{RashbaE}
\end{equation}
The zero overlap between the states in the $\sigma$ bands in Eq.(\ref{hamil_field}) and
Eq.(\ref{hamil_pi_sigma}) imply that only transitions between different
sublattices are allowed.


Defining a $4 \times 4 $ spinor 
\begin{equation} 
\Psi_{K(K')}= \left(
\begin{array}{c}
\Psi_{A \uparrow}(\vec{\bf r})\\
\Psi_{A \downarrow}(\vec{\bf r})\\
\Psi_{B \uparrow}(\vec{\bf r})\\
\Psi_{B \downarrow}(\vec{\bf r})\\
\end{array}
\right)_{K(K')},
\end{equation}

it is possible to join Eqs.(\ref{RashbaCURV}) and (\ref{RashbaE}) in the following compact way:
\begin{equation}
{\cal H}_{\rm R K \pi } =-i \Delta_R \int d^2 \vec{\bf r} \Psi_K^\dag [\hat{\sigma}_+\hat{s}_+ -\hat{\sigma}_-\hat{s}_- ]\Psi_K = \frac{\Delta_{\rm R}}{2} \int d^2 \vec{\bf r} \Psi_K^\dag [\hat{\sigma}_x\hat{s}_y +\hat{\sigma}_y\hat{s}_x ]\Psi_K
\label{Rashba_K}
\end{equation}

\begin{equation}
{\cal H}_{\rm R K' \pi} = -i \Delta_R \int d^2 \vec{\bf r} \Psi_{K'}^\dag [-\hat{\sigma}_+\hat{s}_- +\hat{\sigma}_-\hat{s}_+ ]\Psi_{K'}=\frac{\Delta_{\rm R}}{2} \int d^2 \vec{\bf r} \Psi_{K'}^\dag [\hat{\sigma}_x\hat{s}_y -\hat{\sigma}_y\hat{s}_x ]\Psi_{K'}
\label{Rashba_K'}
\end{equation}
where
\begin{eqnarray}
\Delta_R & = & \Delta_{\cal E}+\Delta_{\rm curv}\nonumber \\ 
\Delta_{\cal E} & =&  \frac{\Delta V_2}{ 2 V_1^2 + V_2^2} \left[\frac{2 \sqrt{2}}{3}  \lambda {e \cal E}\right] \simeq \frac{2 \sqrt{2}}{3} \frac{\Delta \lambda {e \cal E}}{ V_2} \nonumber \\
\Delta_{\rm curv} & = & \frac{\Delta V_1}{ 2 V_1^2 + V_2^2} \left[
( V_{pp \sigma} - V_{pp \pi} ) \left( \frac{a}{R_1} + \frac{a}{R_2} \right)\right]\simeq \frac{\Delta ( V_{pp \sigma} - V_{pp \pi} ) }{V_1}
 \left( \frac{a}{R_1} + \frac{a}{R_2} \right)\left(\frac{V_1}{V_2} \right)^2  .
\label{def_Rashba}
\end{eqnarray}
where the limit $V_1 \ll V_2$ (widely separated $\sigma$ bands) has been considered to approximate the above expressions.

Eqs.(\ref{Rashba_K},\ref{Rashba_K'},\ref{def_Rashba}) constitute one of the most important results of the paper. First, we recover the effective form for the ``Rashba-type'' interaction expected from group-theoretical arguments recently\cite{KM05}. Even more importantly, our result shows that this effective spin-orbit coupling for the $\pi$ bands in graphene to first order in the intraatomic
spin-orbit interaction $\Delta$ is given by two terms:
\begin{itemize} 
\item $\Delta_{\cal E}$: Corresponds to processes due to the intraatomic spin-orbit coupling and the intraatomic Stark effect between different orbitals of the $\pi$ and $\sigma$ bands, together with hopping between neighboring atoms. The mixing between the $\pi$ and $\sigma$ orbitals occurs between the $p_z$ and $s$ atomic orbitals due to the Stark effect $\lambda$ and between the $p_z$ and $p_{x,y}$ due to the atomic spin-orbit coupling $\Delta$. This contribution is the equivalent, for graphene, to the known Rashba spin-orbit interaction\cite{BYCHRASH84} and it vanishes at $\cal E=$0.
\item $\Delta_{\rm curv}$:  Corresponds to processes due to the intraatomic spin-orbit coupling and the local curvature of the graphene surface which couples the  $\pi$ and $\sigma$ bands, together with hopping between neighboring atoms. The mixing between the $\pi$ and $\sigma$ orbitals in this case occurs between $p_z$ and $p_{x,y}$ atomic orbitals both due to the atomic spin-orbit coupling $\Delta$ and due to the curvature. This process is very sensitive to deformations of the lattice along the bond direction between the different atoms where the $p$ part of the the $sp^2$ orbitals is important. 
\end{itemize}

\subsection{Intrinsic spin orbit coupling.}

\begin{figure}
\begin{center}
\includegraphics*[width=6cm]{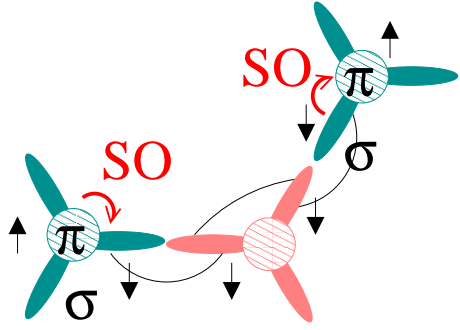}
\caption{``(Color online)'': Sketch of the processes leading to an effective intrinsic term in the $\pi$ band of graphene. Transitions drawn in red (dark grey), and indicated by SO, are mediated by the intraatomic spin-orbit coupling.}
\label{intrinsic_hoppings}
\end{center}
\end{figure}

We can extend the previous analysis to second order in the intraatomic
spin-orbit interaction $\Delta$. We obtain effective couplings between
electrons with parallel spin. The coupling between first nearest neighbors
can be written as:
\begin{equation}
\begin{array}{llclcl}
| p_z 0 \uparrow \rangle &\xrightarrow{\Delta} &| p_x 0 \downarrow
\rangle  &\xrightarrow{V_\sigma} &| p_x 1 \downarrow \rangle
&\xrightarrow{\Delta} | p_z 1 \uparrow \rangle  \\
| p_z 0 \uparrow \rangle  &\xrightarrow{\Delta} &- \frac{1}{2} | p_x 0 \downarrow
\rangle + \frac{\sqrt{3}}{2} | p_y 0 \downarrow \rangle &\xrightarrow{V_\sigma} 
&  \frac{1}{2} |  p_x 2 \downarrow \rangle - \frac{\sqrt{3}}{2} | p_y 2
  \downarrow \rangle &\xrightarrow{\Delta } | p_z 2 \uparrow \rangle 
  \\ | p_z 0 \uparrow \rangle  &\xrightarrow{\Delta} &- \frac{1}{2} | p_x 0 \downarrow
\rangle - \frac{\sqrt{3}}{2} | p_y 0 \downarrow \rangle &\xrightarrow{V_\sigma} &
  \frac{1}{2} |  p_x 3 \downarrow \rangle + \frac{\sqrt{3}}{2} | p_y 3
  \downarrow \rangle &\xrightarrow{\Delta } | p_z 3 \uparrow \rangle \end{array}
\label{hopping_eff_5}
\end{equation}
where the label 0 stands for the central atom.
These three couplings are equal, and give a vanishing contribution at the $K$
and $K'$ points. The intrinsic spin-orbit coupling vanishes for hopping between neighbouring atoms, in agreement with group theoretical arguments\cite{S55, DD65,KM05}. 
We must therefore go to the next order in the hopping. Expanding to next
nearest neighbors, we find a finite contribution to the intrinsic spin-orbit
coupling in a flat graphene sheet, corresponding to processes shown in Fig.[\ref{intrinsic_hoppings}].

In this case, both the dispersive and non dispersive bands contribute to the
effective $\pi - \pi$ coupling, as schematically shown by the different arrows in Fig.[\ref{sigma_Bnds}].
In order to estimate quantatively the magnitude of the intrinsic coupling,
we consider processes represented in Fig.[\ref{intrinsic_hoppings}], which are second order in $\Delta$, 
in momentum space, finally obtaining:
 \begin{eqnarray}
{\cal H}_{{\rm int} K(K')} &= & \pm \frac{3}{4} \frac{\Delta^2}{V_1} 
\frac{V_1^4}{( V_2^2 - V_1^2 ) ( 2 V_1^2 +
  V_2^2 )} \nonumber \\ &\times &\int d^2 \vec{\bf r} \Psi^\dag_{A K(K') \uparrow} ( \vec{\bf r} )
\Psi_{A K(K') \uparrow} ( \vec{\bf r} ) - \Psi^\dag_{A K(K') \downarrow} ( {\bf
  \vec{r}} ) \Psi_{A  K(K') \downarrow} ( \vec{\bf r} ) - \Psi^\dag_{B  K(K')
  \uparrow} ( \vec{\bf r} ) \Psi_{B  K(K') \uparrow} ( \vec{\bf r} ) + 
\Psi_{B K(K') \downarrow} ( \vec{\bf r} ) \Psi_{B K(K') \downarrow} ( \vec{\bf r} ) \nonumber \\
\label{intrinsic}
\end{eqnarray}
where the $\pm$ sign corresponds to $K(K')$ respectively. We define the intrinsic spin-orbit coupling parameter $\Hi$ in the limit $V_1 \ll V_2$ (widely separated
$\sigma$ bands) as:
\begin{equation}
\Hi = \frac{3}{4} \frac{\Delta^2}{V_1} \frac{V_1^4}{( V_2^2 - V_1^2 ) ( 2 V_1^2 +
  V_2^2 )} \simeq \frac{3}{4} \frac{\Delta^2}{V_1} \left( \frac{V_1}{V_2} \right) ^4 
\end{equation}
Our Hamiltonian, Eq.(\ref{intrinsic}), is equivalent to the one derived in \cite{KM05} 

\begin{equation}
{\cal H}_{\rm T SO \, \, \, intrinsic } = \int d^2 \vec{\bf r} \Hi \Psi^\dag [\hat{\tau}_z \hat{\sigma}_z\hat{s}_z ]\Psi
\label{Intrinsic_K'}
\end{equation}
where $\hat{\tau}_z = \pm 1$ denotes the K($K'$) Dirac point and $ \Psi= \left(  \Psi_{K},\Psi_{K'}\right )^T$.


\section{Numerical estimates.}
We must now estimate $\Hi$, $\Delta_{\cal E}$ and $\Delta_{\rm curv}$. We have $\lambda= 3 a_o/Z\approx0.264$\AA\cite{BJ83}, where $Z=6$ for carbon and $a_o$ is the Bohr radius, ${\cal E} \approx$ 50V/300nm\cite{KM05,Netal04}, the atomic spin-orbit splitting for carbon $\Delta = 12$meV $\rightarrow 1.3 \times 10^{2}$K\cite{SCR00,HS63}, the energy difference between the $\pi$-2p$_z$ orbitals and the  $\sigma$-sp$^2$ orbitals $\epsilon_\pi-\epsilon_\sigma \sim  (14.26-11.79)$eV$=2.47$eV, the energy difference between the 2p and the 2s atomic orbitals $\epsilon_s - \epsilon_p \sim (19.20-11.79)$eV$=7.41$eV and the hoppings between the $2s, 2p_x,2p_y,2p_z$ orbitals of neighbouring atoms as $V_{sp \sigma} \sim 4.2$eV, $V_{ss \sigma} \sim -3.63$eV and $V_{pp \sigma} \sim 5.38$eV and $V_{pp \pi} \sim -2.24$eV \cite{TL88,TS91}.  We have $V_1=2.47$eV and $V_2=6.33$eV. We estimate $\Hi\simeq (3\Delta^2/4 V_1) (V_1/V_2)^4 \sim 0.1 \times 10^{-5}$ eV $\rightarrow$ 0.01K. $\Hi$ is two orders of magnitude smaller than the estimate in\cite{KM05}. The discrepancy seems to arise first, because the intrinsic spin-orbit splitting $\Hi$  estimated here is proportional to the square of the intraatomic spin-orbit coupling $\Delta^2$, instead of being proportional to it, as roughly estimated in\cite{KM05}. Besides, a detailed description of the $\sigma$ bands is necessary to obtain the correct estimate. Spin-orbit splittings of order 1-2.5 K have been discussed in the literature for graphite\cite{BCP88}. However in graphite, the coupling between layers is important and influences the effective value of the spin-orbit splitting, typically being enhanced with respect to the single layer value\cite{DD65,MY62,Y63}.

For the other two couplings, we use the full expression obtained for $\Delta_{\cal E }$ and $\Delta_{\rm curv}$ and not the limiting form $V_1 \ll V_2$, in order to be as accurate as possible. First, we obtain $\Delta_{\cal E}=\left( 2 \sqrt{2}/3\right) \left(\lambda e {\cal E } \Delta V_2/\left(2V_1^2 +V_2^2\right) \right) \sim 0.6 \times 10^{-5}$ eV $\rightarrow$ 0.07K. This estimate for  $\Delta_{\cal E }$, depends on the external electric field chosen. Our estimate, for the same value of the electric field, is two orders of magnitude bigger than the estimate in\cite{KM05}. So far curvature effects have been excluded. Curvature effects will increase the total value for the effective  spin-orbit interaction in graphene. Graphene samples seem to have an undulating surface \cite{Metal06}. The ripples observed seem to be several \AA$~$height and a few tens nm laterally\cite{Metal06}. First we consider the simplest example of a ripple being a half-sphere of radius R. The part of the sphere which intersects the plane of flat graphene and therefore constitutes the ripple, is assumed to have a typical height $h \ll R$ so the radius R is roughly of the same order of magnitude of the lateral size in this case. It seems possible to identify  ripples of lateral size ranging  ~ 50nm -100nm in \cite{Metal06}. Choosing $a=1.42$\AA$~$ and $ R_1\sim R_2 \sim 50$nm\cite{Metal06}, we obtain $\Delta_{\rm curv}= \left( 2a/R\right) \left(\left(V_{pp \sigma} -V_{pp \pi} \right) \Delta V_1/\left(2V_1^2 +V_2^2\right) \right)\sim 2.45 \times 10^{-5}$eV $\rightarrow$ 0.28 K. Choosing $ R_1\sim R_2 \sim 100$nm we obtain $\Delta_{\rm curv} \sim 1.22 \times 10^{-5}$eV $\rightarrow$ 0.14 K. Now we consider a different model where we assume that the sample has random corrugations of height $h$ and length $l$\cite{MG06}. The graphene surface presents then an undulating pattern of ripples of average radius $R \sim l^2/h$\cite{MG06}. Choosing $l \sim 100$\AA $~$ and $h \sim 10$ \AA\cite{MG06}, we obtain again $R \sim 100$nm, which leads to the same value for $\Delta_{\rm curv} \sim 1.22 \times 10^{-5}$eV $\rightarrow$ 0.14 K. In any case, it seems clear that due to curvature effects, the effective spin-orbit coupling in graphene could be higher for curved graphene than for perfectly flat graphene. Moreover, spin-orbit coupling in (curved) graphene would be present even for $\cal E=$0. A more detailed discussion/study of the local curvature/corrugation of graphene is needed in order to obtain more accurate estimates. 

To conclude this section we present the effective hamiltonian for the 
$\pi$-bands of graphene including the spin-orbit interaction:
\begin{table}
\begin{tabular}{||c|c|c||}
\hline \hline
Intrinsic coupling: $\Hi$ & $\frac{3}{4} \frac{\Delta^2}{V_1} \left( \frac{V_1}{V_2} \right)^4$ &  0.01K\\ \hline
Rashba coupling (electric field ${\cal E}\approx 50V/300$nm): $\Delta_{\cal E}$ & $\frac{2 \sqrt{2}}{3} \frac{\Delta \lambda {e \cal E}}{ V_2}$& 0.07K\\ \hline
Curvature coupling: $\Delta_{\rm curv}$ & $\frac{\Delta ( V_{pp \sigma} - V_{pp \pi} ) }{V_1}
 \left( \frac{a}{R_1} + \frac{a}{R_2} \right)\left(\frac{V_1}{V_2} \right)^2  $&  0.2K\\ \hline
\end{tabular}
\caption{Dependence on band structure parameters, curvature, and electric  field of the spin orbit couplings discussed in the text in the limit $V_1 \ll V_2$ (widely separated
$\sigma$ bands). The parameters used are $\lambda \approx0.264$\AA\cite{BJ83}, ${\cal E} \approx$ 50V/300nm\cite{KM05,Netal04}, $\Delta = 12$meV
\cite{SCR00,HS63},$V_{sp \sigma} \sim 4.2$eV, $V_{ss \sigma} \sim -3.63$eV, $V_{pp \sigma} \sim 5.38$eV and $V_{pp \pi} \sim -2.24$eV \cite{TL88,TS91}, $V_1=2.47$eV, $V_2=6.33$eV, $a=1.42$\AA,$~$ $l \sim 100$ \AA, $h \sim 10$ \AA $~$ and $R \sim 50-100$nm.} 
\label{numerical}
\end{table}

\begin{equation}
{\cal H}_{T} = \int d^2 \vec{\bf r}  \Psi^\dag \left( -i \hbar v_F [\hat{\sigma}_x \hat{\partial}_x + \hat{\tau}_z \hat{\sigma}_y \hat{\partial}_y] + \Hi [\hat{\tau}_z \hat{\sigma}_z\hat{s}_z ] + \frac{\Delta_R }{2}[ \hat{\sigma}_x\hat{s}_y + \hat{\tau}_z \hat{\sigma}_y\hat{s}_x] \right) \Psi
\label{Effective_H_SO}
\end{equation}
where $\hbar v_F =\sqrt{3} \gamma_o $a$ /2 $, a$~ \sim 2.46$ \AA $~$being the
lattice constant for graphene, $\gamma_o \sim 3$eV the McClure intralayer
coupling constant\cite{M57,DD65}, $\Hi\sim
0.01$K, $\Delta_R =\Delta_{\cal E} +\Delta_{\rm curv}$ the Rashba-Curvature coupling (RCC), where $\Delta_{\cal E} \sim 0.07$K for ${\cal E} \approx$ 50V/300nm and $\Delta_{\rm curv} \sim 0.2$K. Table[\ref{numerical}] summarizes the main results obtained in the paper for the effective spin-orbit couplings in a graphene layer.


\section{Application to fullerenes, nanotubes and fullerene caps.}
\subsection{Spherical fullerenes.}
When topological deformations in the form of pentagons are introduced in the hexagonal lattice of graphene, curved structures form.
If 12 pentagons are introduced, the graphene sheet will close itself into a sphere forming the well know fullerene structure \cite{KROTO85}. The usual hexagonal lattice ``lives'' now on a sphere and presents topological defects in the form of pentagons. The continuum approximation to a spherical fullerene leads to two decoupled
 Dirac equations in the presence of a fictitious monopole of charge $\pm
 3/2$ in the center of the sphere which accounts for the presence of the 12 pentagons\cite{GGV92,GGV93b}.  
The states closest to the Fermi level are four triplets at $\epsilon = 0$.  We consider the effect of the spin-orbit
 coupling on these triplets first.
  
Both the coupling induced by the Rashba-Curvature, 
Eq.(\ref{Rashba_K}, \ref{Rashba_K'}), and the intrinsic coupling,
Eq.(\ref{intrinsic}), are written in a local basis of
wavefunctions where the spin is oriented perpendicular to the graphene
sheet, $| \theta , \phi , \perp \uparrow \rangle , | \theta , \phi , \perp
\downarrow \rangle$. It is useful to relate this local basis with a fixed basis independent of the curvature. Such relation depends on the curvature of the graphene sheet considered. In the case of a fullerene the graphene sheet is in a sphere. The details are given in Appendix B.

The gauge field associated to the presence of fivefold rings in the fullerene
can be diagonalized using the basis:
\begin{eqnarray}
\tilde{\Psi} _{A {\cal K} {\bf \vec{k}} s} ( \vec{\bf r} )  &= 
&\Psi_{A K {\bf \vec{k}} s} ( \vec{\bf r} ) +
i \Psi_{B K' {\bf \vec{k}} s} ( \vec{\bf r} ) \nonumber \\
\tilde{\Psi}_{B {\cal K'} {\bf \vec{k}} s} ( \vec{\bf r} ) &= -i &\Psi_{B K' {\bf
  \vec{k}} s} ( \vec{\bf r} ) + \Psi_{A K {\bf \vec{k}} s} ( {\bf
\vec{r}} ).
\label{transformation1}
\end{eqnarray}
Equivalent transformation is obtained exchanging $A \leftrightarrow B$.

 In this basis, the
 wavefunctions of the zero energy states are\cite{GGV92}:
\begin{eqnarray}
| +1 \, s \, {\cal K} \rangle &\equiv &\sqrt{\frac{3}{4 \pi}} \cos^2 \left(
 \frac{\theta}{2} \right) e^{i \phi } \left(  \begin{array}{c}
| A K \rangle  \\ i  | B K' \rangle \end{array} \right)  \otimes | s
\rangle \nonumber \\
| 0 \, s \, {\cal K} \rangle &\equiv &\sqrt{\frac{3}{2 \pi}} \sin \left(
 \frac{\theta}{2} \right) \cos \left(  \frac{\theta}{2} \right)  \left(  \begin{array}{c}
| A K \rangle  \\ i  | B K' \rangle \end{array} \right)  \otimes | s
\rangle \nonumber \\
| -1 \, s \, {\cal K} \rangle &\equiv &\sqrt{\frac{3}{4 \pi}} \sin^2 \left(
 \frac{\theta}{2} \right) e^{- i \phi } \left(  \begin{array}{c}
| A K \rangle  \\ i  | B K' \rangle \end{array} \right)  \otimes | s
\rangle \nonumber \\
| + 1 \, s \, {\cal K}' \rangle &\equiv &\sqrt{\frac{3}{4 \pi}} \sin^2 \left(
 \frac{\theta}{2} \right) e^{i \phi } \left(  \begin{array}{c}
| A K \rangle  \\ - i  |B K' \rangle \end{array} \right)  \otimes | s
\rangle \nonumber \\
| 0 \, s \, {\cal K}' \rangle &\equiv &- \sqrt{\frac{3}{2 \pi}} \sin \left(
 \frac{\theta}{2} \right) \cos \left(  \frac{\theta}{2} \right)  \left(  \begin{array}{c}
| A K \rangle  \\ - i  |B K' \rangle \end{array} \right)  \otimes | s
\rangle \nonumber \\
| -1 \, s \, {\cal K}' \rangle &\equiv &\sqrt{\frac{3}{4 \pi}} \cos^2 \left(
 \frac{\theta}{2} \right) e^{- i \phi } \left(  \begin{array}{c}
| A K \rangle  \\ - i  |B K' \rangle \end{array} \right)  \otimes | s
\rangle
\label{wavefunctions}
\end{eqnarray}
where $|A K \rangle$ and $| B K' \rangle$ are envelope functions associated to
the $K$ and $K'$ points of the Brillouin zone and corresponding to states located at the $A$  and $B$ sublattices respectively. Note that, at zero energy, states at $K(K')$ are only located at sublattice $A(B)$ sites. $| s
\rangle$ denotes the usual spinor part of the wave function corresponding to the electronic spin $s= \uparrow, \downarrow$.
The  hamiltonian ${\cal H}_{\rm int}$ couples orbitals in the same sublattice  whereas  
${\cal H}_{\rm  R}$ couples orbitals in different sublattices. So ${\cal H}_{\rm R}$ has zero matrix elements between zero energy states, as it does not induce intervalley scattering, mixing $K$ and $K'$ states \cite{MG06}.  

In the $\{ |+1 \uparrow \rangle ,|+1 \downarrow \rangle,|0 \uparrow \rangle ,|0 \downarrow \rangle,|-1 \uparrow \rangle ,|-1 \downarrow \rangle \}$ basis, the Hamiltonian for the ${\cal K}$ point of a fullerene looks like:
\begin{equation}
{\cal H}^{{\cal K}}_{\rm S-O \, int} = \left( \begin{array}{cccccc} 
    \Hi & 0 &  0 & 0 &0 &  0\\ 
     0 & -\Hi& \sqrt{2} \Hi &0 & 0& 0 \\ 
     0 & \sqrt{2} \Hi &0& 0 &0 & 0 \\
     0& 0 & 0& 0 &\sqrt{2} \Hi & 0 \\
     0& 0 & 0& \sqrt{2} \Hi & -\Hi& 0 \\
     0& 0 & 0& 0 &0 &  \Hi\\
    \end{array} \right)
 \label{H_fullerenes_K}
\end{equation}
The Hamiltonian for ${\cal K}'$ is ${\cal H}^{{\cal K}'}_{\rm S-O \, int}=-{\cal H}^{{\cal K}}_{\rm S-O \, int}$.

Diagonalizing the Hamiltonian Eq. (\ref{H_fullerenes_K}), we obtain that each set of spin degenerate triplets obtained in the
absence of the spin-orbit interaction split into:
\begin{eqnarray}
\epsilon &= & +\Hi\nonumber \rightarrow \Psi_{\Delta}: \{ | +1 \uparrow \rangle, | -1 \downarrow \rangle, \sqrt{\frac{1}{3}} |+1 \downarrow \rangle + \sqrt{\frac{2}{3}} |0 \uparrow \rangle, \sqrt{\frac{1}{3}} |-1 \uparrow \rangle + \sqrt{\frac{2}{3}} |0 \downarrow \rangle\}  \nonumber \\
\epsilon &= &-2 \Hi \rightarrow \Psi_{- 2 \Delta}: \{\sqrt{\frac{2}{3}} |+1 \downarrow \rangle - \sqrt{\frac{1}{3}} |0 \uparrow \rangle, \sqrt{\frac{2}{3}} |-1 \uparrow \rangle - \sqrt{\frac{1}{3}} |0 \downarrow \rangle \}
\end{eqnarray}

Each of these solutions is doubly degenerate, corresponding to the ${\cal K}$ and ${\cal K}'$ points.
In principle, many body effects associated to the electrostatic interaction
can be included by following the calculation discussed in\cite{GGV93c}.


\subsection{Spin-orbit coupling in nanotubes.}
The previous continuum analysis can be extended to nanotubes. We use cylindrical
coordinates, $z , \phi$, and, as before, define the spin orientations $|
\uparrow \rangle , | \downarrow \rangle$ as parallel and antiparallel to the
$z$ axis. The matrix elements relevant for this geometry can be easily obtained
from Eq.(\ref{local_spin}) in Appendix B, by choosing $\theta =\pi/2$.
 The eigenstates of the nanotube can be classified by longitudinal momentum,
$k$, and by their angular momentum $n$, $\epsilon_{\pm , k , n} = \pm \hbar \vf \sqrt{k^2 + n^2 / R^2}$, where $R$ is the
radius of the nanotube. After integrationg over the circumference of the nanotube $\int d\phi$, the Hamiltonian of a nanotube 
including spin-orbit interaction is:
\begin{equation}
{\cal H}_{\rm S-O \, R} \left( \begin{array}{c}
|A \tau \rangle \\
|B  \tau \rangle \\
\end{array} \right) = \left( \begin{array}{cc} 
    0 &  \hbar \vf (k-in/R) +  \tau ~ i \Delta_R \pi \hat{s}_z\\ 
    \hbar \vf (k+in/R) -  \tau ~ i \Delta_R \pi \hat{s}_z & 0\\
    \end{array} \right)
\left( \begin{array}{c}
|A  \tau \rangle \\
|B  \tau \rangle \\
\end{array}
\right)
 \label{H_carbonanotubesK}
\end{equation}
where the $\tau = \pm 1$ corresponds to the $K(K')$ Dirac point. Note the basis states 
$ | A \tau \rangle$ and $ | B  \tau \rangle$  used to define Eq.(\ref{H_carbonanotubesK}) are spinors in spin subspace where the matrix $\hat{s}_z$ acts on (see Appendix C for details). The contribution from the intrinsic spin-orbit $\Hi$ becomes zero after integrating over the nanotube circumference (Appendix C). 
The spin-orbit term $ i \Delta_R \pi \hat{s}_z$ in  Eq.(\ref{H_carbonanotubesK}) is equivalent to the term proportional to $\hat{\sigma}_y$ obtained in Eq. (3.15, 3.16) of Ref. \cite{A00}. It is important to note that the spin orientations $|\uparrow \rangle , | \downarrow \rangle$ in Eq.(\ref{H_carbonanotubesK}) are defined along the nanotube axis, whereas the spin orientations used in Eq. (3.15, 3.16) of Ref. \cite{A00} are defined perpendicular to the nanotube surface. On the other hand, we do not find any contribution similar to the term proportional to $\sigma_x(r)$ in Eq.(3.15, 3.16) in Ref. \cite{A00}. In any case, such contributions are not important as they vanish after integrating over the circumference of the nanotube\cite{A00}. Besides, our results are in agreement with the results obtained in \cite{Metal02}.

The energies near the Fermi level, $n=0$, are changed
by the spin-orbit coupling, and we obtain:
\begin{equation}
\epsilon_{k} = \pm \sqrt{\left( \pi \Delta_{\rm R} \right)^2 + (\hbar \vf k )^2}.
\end{equation} 
There is an energy gap $\pi \Delta_{\rm R}$ at low energies, in agreement with the results in\cite{A00,CLM04}. The $\pi \Delta_{\rm R}$ gap originates as a consequence of the Berry phase gained by the electron and hole quasiparticles after completing a closed trajectory around the circumference of the nanotube under the effect of spin-orbit interaction  $\Delta_{\rm R}$ \cite{A00}. Similarly, $\Delta_{\rm R}$ will give rise to a small spin splitting for $n \neq 0$ \cite{A00,CLM04}
\begin{equation}
\epsilon_{k} = \pm \sqrt{\left( \pi \Delta_{\rm R} \right)^2 + (\hbar \vf )^2 (k^2+(n/R)^2)+2 (n/R) \hbar \vf \Delta_R \pi \hat{s}_z  }.
\end{equation}

For a single wall nanotube of radius $R_1 \sim 6, 12, 24$\AA $~$ and $R_2 \rightarrow \infty$, $a \sim 1.42$\AA$~$ and for $\cal E=$0, we get $\Delta_R \sim 12,6,3$K respectively.

\subsection{Nanotube caps.}
\subsubsection{Localized states at zero energy.}
\begin{figure}
\begin{center}
\includegraphics*[width=6cm]{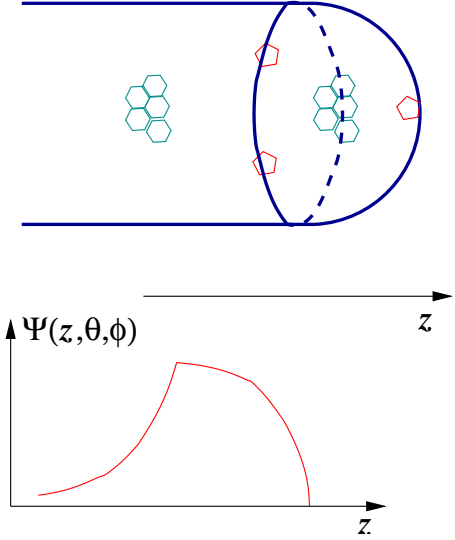} 
\caption{``(Color online)'': (Up) Sketch of the matching scheme used to build a 
zero energy state at a fullerene cap. (Down)The wavefunction is one half of a zero energy state at the cap, matched to a decaying state towards the bulk of the nanotube. See text for details.}
\label{cap_fig_2}
\end{center}
\end{figure}

 An armchair $( 5 N \times 5 N )$ nanotube can be ended by a spherical fullerene cap. The cap
contains six pentagons and $5 N ( N + 1)/2 - 5$ hexagons. When $N = 3 \times M$,
the nanotube is metallic. An example of such a fullerene cap is
given in Fig.[\ref{cap_fig}].
\begin{figure}

\begin{center}
\includegraphics*[width=4cm]{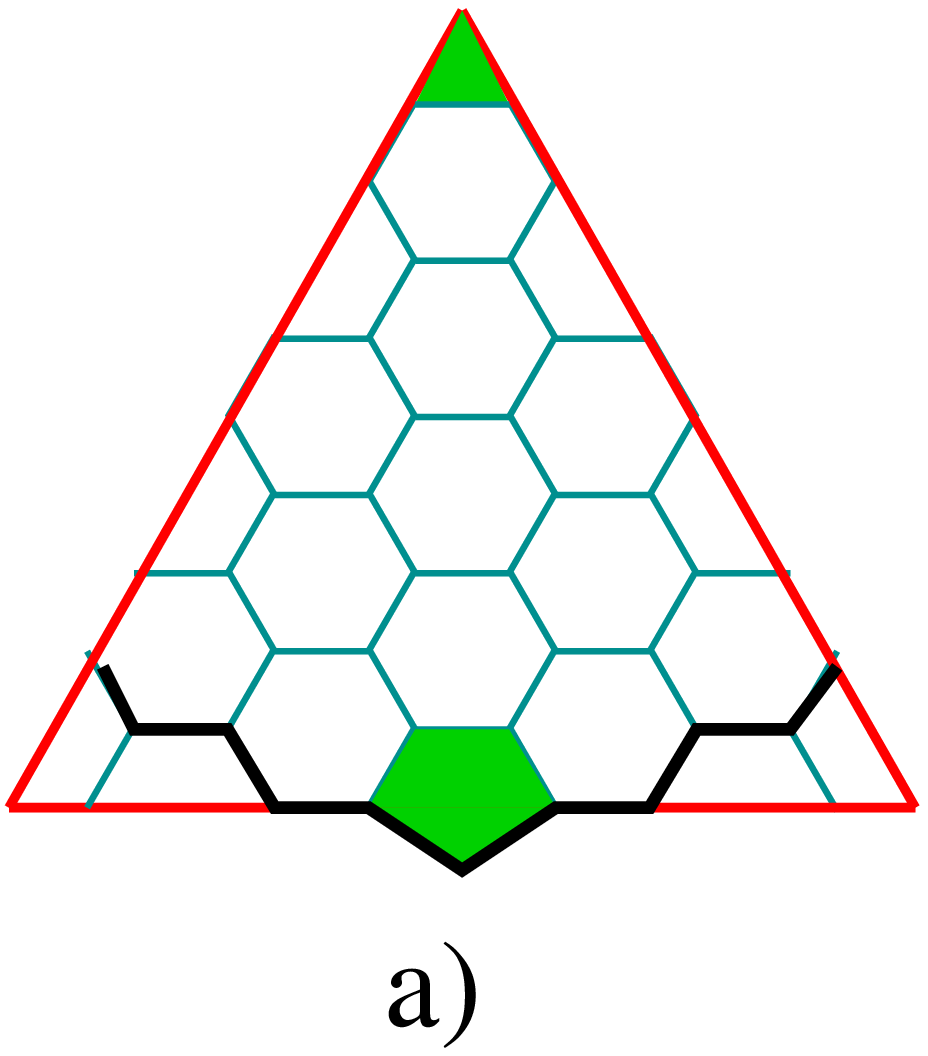} 
\label{cap_fig}
\includegraphics*[width=7cm]{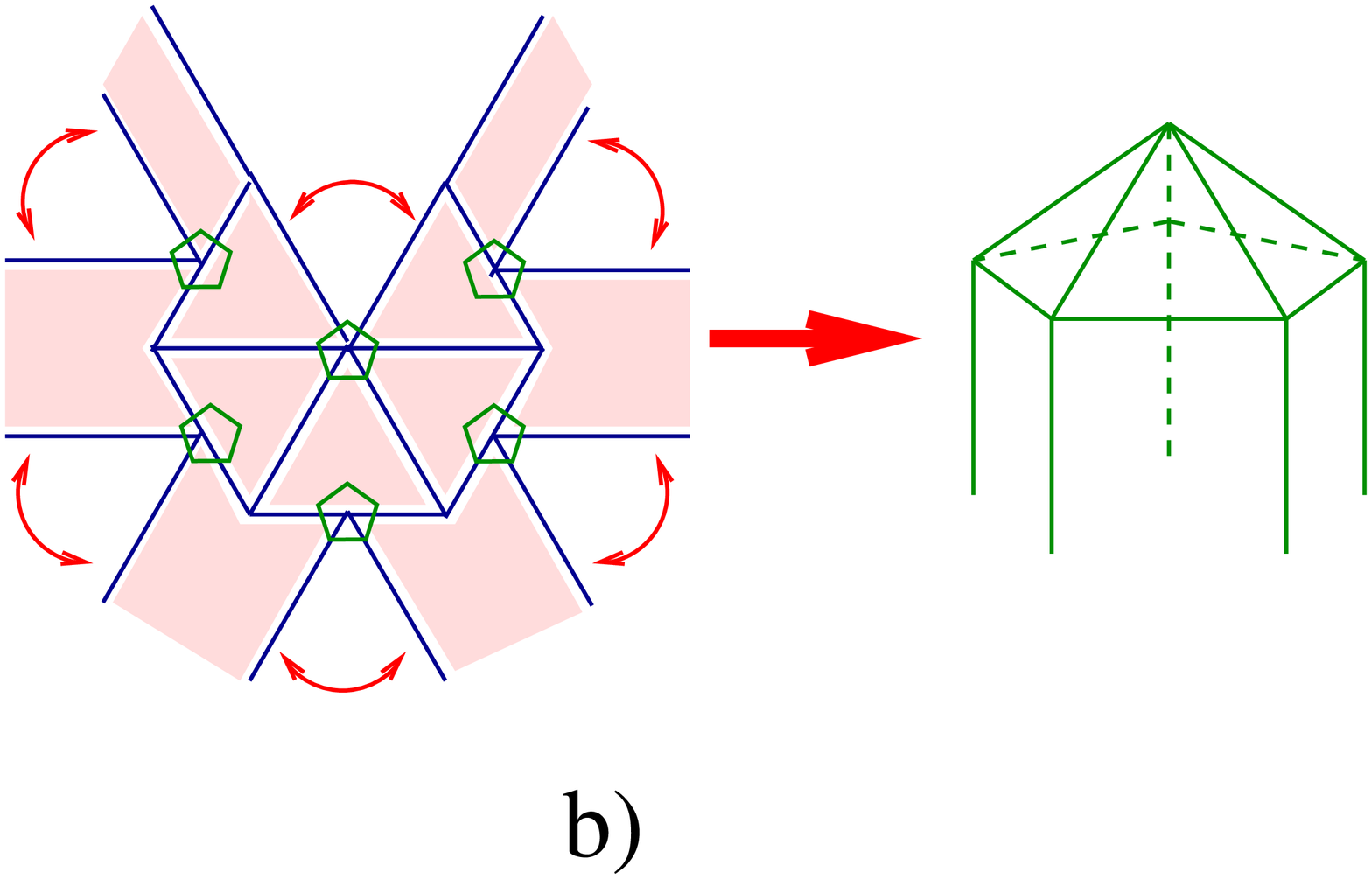}
\caption{``(Color online)'': Left: One fifth of a fullerene cap closing an armchair nanotube. The full cap is obtained by gluing five triangles like the one in the figure together, forming a pyramid. A pentagon is formed at the apex of the
  pyramid, from the five triangles like the one shaded in green (grey) in the
  figure. The edges of the cap are given by the thick black line. The cap
  contains six pentagons and seventy hexagons, and it closes a $25 \times 25$
  armchair nanotube. Right: Sketch of the folding procedure of a flat
  honeycomb lattice needed to obtain an armchair nanotube capped by a semispherical fullerene.}
\label{cap_fig}
\end{center}
\end{figure}
The boundary between the semispherical fullerene and the nanotube is a
circle (Fig.[\ref{cap_fig_2}]). The solutions of the continuum equations have to be continuous accross
this boundary, and they have to satisfy the Dirac equations appropriate for
the sphere in the cap and for the torus in the nanotube respectively.

The boundary of the nanotube in the geometry shown in Fig.[\ref{cap_fig_2}] is a
zigzag edge. Hence, zero energy states can be defined\cite{WS00,W01}, 
which at this boundary will have a finite amplitude on one sublattice and zero on
the other. There is a zero energy state, $| \Psi_n \rangle $
at this boundary, for each value of the angular momentum around the nanotube $n$. 
They decay towards the bulk of the nanotube as 
\begin{eqnarray}
\Psi_n ( z , \phi , K ) &= &C e^{i n \phi} e^{- ( n z ) / R} \, \, \, \, \,
\, \, n > 0 \nonumber \\
\Psi_n ( z , \phi , K' ) &= &C e^{i n \phi} e^{- ( n z ) / R} \, \, \, \, \,
\, \, n <  0
\label{boundary_states}
\end{eqnarray}
where we are assuming that the nanotube is in the half space $z>0$ (see Fig.[\ref{cap_fig_2}]).

A zero energy state in the whole system can be defined if there are states inside
the gap of the nanotube $\pi \Delta_{\rm R}$, which can be matched to the states defined in
Eq.(\ref{boundary_states}). At the boundary we have $\theta = \pi / 2 , ~ \cos
( \theta / 2 ) = ~\sin ( \theta / 2 ) = 1/\sqrt{2}$. Hence, we can  combine states $|
l \, s \, {\cal K} \rangle$ and $| l \, s \, {\cal K}' \rangle , l = \pm 1$ in
Eq.(\ref{wavefunctions}) in such a way that the amplitude at the boundary on
a given sublattice vanishes:  

\begin{eqnarray}
| +1 \, s \rangle_A \equiv \frac {1}{\sqrt{2}} \left( | +1 \, s \, {\cal K} \rangle + | + 1 \, s \, {\cal K}' \rangle \right) =
 \sqrt{\frac{3}{8 \pi}} e^{i \phi } \left(  \begin{array}{c}
 | A K \rangle  \\ 0 \end{array} \right)  \otimes | s\rangle \nonumber\\
| -1 \, s \rangle_A \equiv \frac {1}{\sqrt{2}} \left( | -1 \, s \, {\cal K} \rangle + | - 1 \, s \, {\cal K}' \rangle \right) =
\sqrt{\frac{3}{8 \pi}} e^{-i \phi } \left(  \begin{array}{c}
 | A K \rangle  \\ 0 \end{array} \right)  \otimes | s\rangle\nonumber\\\nonumber\\
| +1 \, s \rangle_B \equiv \frac {1}{\sqrt{2}} \left( | +1 \, s \, {\cal K} \rangle - | + 1 \, s \, {\cal K}' \rangle \right) =
 \sqrt{\frac{3}{8 \pi}} e^{i \phi } \left(  \begin{array}{c}
 0  \\ i | B K'\rangle \end{array} \right)  \otimes | s\rangle \nonumber\\
| -1 \, s \rangle_B \equiv \frac {1}{\sqrt{2}} \left( | -1 \, s \, {\cal K} \rangle - | - 1 \, s \, {\cal K}' \rangle \right) =
\sqrt{\frac{3}{8 \pi}} e^{-i \phi } \left(  \begin{array}{c}
  0 \\ i | B K'\rangle \end{array} \right)  \otimes | s\rangle
\label{matching}
\end{eqnarray}

These combinations match the states decaying
into the nanotube, Eq.(\ref{boundary_states}). This fixes the constant $C$ in Eq.(\ref{boundary_states}) 
to be $C=\sqrt{3/(8 \pi)}$.
  Note that the wavefunctions with $l=0$ can only
be matched to states that not decay into the bulk of the nanotube, \emph{i.e.} with $n=0$.

Thus, there are two states per spin $s$, $| +1 \, s \rangle, | -1 \, s \rangle$, localized
at the cap and with finite chirality, $n = \pm 1$.
A sketch of this procedure is shown in Fig.[\ref{cap_fig_2}].
 This continuum approximation is in general agreement with
the results in\cite{YA01}.

As in the case of a spherical fullerene, only the intrinsic spin-orbit
coupling mixes these states. The energies of the $| \pm 1 \, s \rangle_A$ states are not affected by $\Hi$, as the contribution from $| +1 \, s \rangle,$ is cancelled by the contribution from $| -1 \, s \rangle$. On the other hand, the states $| \pm 1 \, s \rangle_B$ split in energy, as the $| +1 \, s \rangle$ and $ | -1 \, s \rangle$ contributions add-up: 
\begin{eqnarray}
| + 1 , s=\uparrow,\downarrow \rangle_B \rightarrow \epsilon_{\uparrow,\downarrow}=  \pm \Hi \nonumber\\
| - 1 , s=\uparrow,\downarrow \rangle_B \rightarrow \epsilon_{\uparrow,\downarrow}=  \mp \Hi. 
\end{eqnarray}
Note that each state has a finite chirality.

\subsubsection{Localized states induced by the spin-orbit interaction.}
The remaining states of a spherical fullerenes have multiplicity $2l+1$,
where  $l \ge 2$ and energy $\epsilon_l = \pm \hbar \vf / R \sqrt{l (l+1)- 2}$. The
angular momentum of these states along a given axis, $m$, is $-l \le m \le l$.
The subbands of the nanotube with angular momentum $\pm m$, have  gaps within
the energy interval
$- \Delta_m = - \hbar \vf | m | / R \le \epsilon \le \Delta_m = \vf | m | / R
$. Thus, there is a fullerene eigenstate with $l=2$ and angular momentum
$m=\pm 2$ which lies at the gap edge of the nanotube subbands with the same
momentum. The fullerene state is:
\begin{equation}
\Psi_{l=2 \, m=2} ( \theta , \phi ) \equiv \frac{1}{4 \sqrt{2 \pi}} e^{2 i
  \phi}
\left(  \begin{array}{c} \sin ( \theta ) \left[ 1 + \cos ( \theta ) \right] | K
  \rangle - [ 1 + \cos ( \theta ) ]^2 | K' \rangle \\ i [  1 + \cos ( \theta )
  ]^2 | K \rangle + i  \sin ( \theta ) \left[ 1 + \cos ( \theta ) \right] | K'
  \rangle \end{array} \right) 
\end{equation}
which can be matched, at $\theta = \pi/2$, to the nanotube eigenstate:
\begin{equation}
\Psi_{m=2} ( z , \phi ) \equiv \frac{1}{4 \sqrt{2 \pi}} e^{2 i
  \phi} \left( \begin{array}{c} | K \rangle - | K' \rangle \\ i | K \rangle +
  i | K' \rangle \end{array} \right)
\label{wf_cap_2}
\end{equation}
The spin orbit coupling acts as a position dependent potential on this state,
and it shifts its energy into the $m=2$ subgap, leading to the formation of another
localized state near the cap. 

In the following, we consider only the Rashba-Curvature
coupling $\Delta_R$. In order to analyze the extension of the state, we assume that
the localized state decays in the nanotube, $z>0$ as:
\begin{equation}
\Psi_{m=2} ( z , \phi ) \equiv \frac{C}{4 \sqrt{2 \pi}} e^{2 i
  \phi} e^{\kappa z / R}
\left( \begin{array}{c} | K \rangle - | K' \rangle \\ i | K \rangle +
  i | K' \rangle \end{array} \right)
\label{state_3}
\end{equation}
We match this wavefunction to that in Eq.(\ref{wf_cap_2}) multiplied by the
same normalization constant, $C$. We
assume that the state is weakly localized below the band edge, so that the main part of the
wavefunction is in the nanotube, and $\kappa \ll 1$. Then, we can neglect the
change in the spinorial part of the wavefunction, and we will fix the
relative components of the two spinors as in Eq.(\ref{wf_cap_2}) and Eq.(\ref{state_3}).
The normalization of the total wavefunction implies that:
\begin{equation}
C^{-2} = \frac{13}{16} + \frac{1}{4 \kappa}
\end{equation}
where the first term in the r.h.s. comes from the part of the wavefunction
inside the cap, Eq.(\ref{wf_cap_2}), and the second term is due to the weight
of the wavefunction inside the nanotube, Eq.(\ref{state_3}). As expected, 
 when the state becomes delocalized, $\kappa
\rightarrow 0$, the main contribution to the normalization arises from the ``bulk'' part of the wavefunction.


The value of $\kappa$ is fixed by the energy of the state:
\begin{equation}
\kappa^2 = n^2 - \frac{\epsilon^2 R^2}{(\hbar\vf)^2}
\label{kappa}
\end{equation}
where the energy of the state $\epsilon$ is inside the subgap $\Delta_m$ of the nanotube
because of the shift induced by the spin orbit interaction $\Delta_R$. 

We now calculate the contribution to the energy of this state from the Rashba-Curvature
coupling, which is now finite, as this state has weight on the two sublattices:
\begin{equation}
\epsilon_{\rm Rashba} \approx    \pm C^2 \Delta_{\rm R} \left( \frac{1}{16 \kappa} +
  \frac{31}{80} \right) \approx \pm \frac{\Delta_{\rm R}}{4} \left( 1 - \frac{59
  \kappa}{20} \right)
\label{Rashba_cap}
\end{equation}
The r.h.s. in Eq.(\ref{Rashba_cap}) can be described as the sum of a bulk
term, $\pm \Delta_{\rm R} / 4$, and a term due to the presence of the cap,
whose weight vanishes as the state becomes delocalized, $\kappa \rightarrow
0$. The absolute value of the Rashba-Curvature contribution is reduced with respect to
the bulk energy shift, which implies that the interaction is weaker at the
cap.

The bulk nanotube bands are split into two spin subbands which are shifted in
opposite directions. The surface states analyzed here are shifted by a
smaller amount, so that the state associated to the subband whose gap
increases does not overlap with the nanotube continuum. Combining the
estimate of the energy between the gap edge and the surface state in
Eq.(\ref{Rashba_cap}) and the constraint for $\kappa$ in  Eq.(\ref{kappa}),
we find:
\begin{equation}
\kappa \approx \frac{59 \Delta_{\rm R} R}{20 \hbar \vf}
\end{equation} 
Finally, the separation between the energy of the state and the subgap edge is:
\begin{equation}
\epsilon \approx \frac{\hbar \vf \kappa^2}{8 R}
\end{equation}
For a C$_{60}$ fullerene of radius $R\sim 3.55$\AA $~$ we obtain, for $\cal E=$0, a value $\Delta_R/4 \sim 3$K.
 This effect of the spin orbit interaction 
can be greatly enhanced in nanotube caps in an external electric
field, such as those used for field emission devices\cite{FECAPS05}. In this case, 
spin-orbit interaction in may allow for spin dependent field emission of such devices. 
The applied field also
modifies the one electron states, and a detailed analysis of this situation
lies outside the scope of this paper.
\section{Conclusions.}
We have analyzed the spin orbit interaction in graphene and similar
materials, like nanotubes and fullerenes. We have extended previous
approaches in order to describe the effect of the intraatomic spin orbit
interaction on the conduction $\pi$  and valence $\sigma$ bands.  Our scheme allows us
to analyze, on the same footing, the effects of curvature and perpendicular applied
electric field. Moreover, we are able to obtain realistic
estimates for the intrinsic $\Hi$ and Rashba-Curvature $\Delta_R= \Delta_{\cal E}+\Delta_{\rm curv}$ effective spin-orbit couplings in graphene. We have shown that spin-orbit coupling for flat graphene is rather weak $\Hi \sim 10$mK and $\Delta_{\cal E} \sim  70$mK for $\cal E$=$50$V$/300$nm. Moreover curvature at the scale of the distance between neighbouring atoms increases the value of the spin orbit coupling in graphene $\Delta_{\rm curv} \gtrsim \Delta_{\cal E} \gg \Hi$. This is because local curvature mixes the $\pi$ and $\sigma$ bands. Graphene samples seem to have an undulating surface \cite{Metal06}. Our estimate for the typical observed ripples indicates that $\Delta_{\rm curv}$ could be of order $\sim 0.2K$.  A more detailed study of the curvature of graphene samples is needed in order to obtain a more precise estimate. 
We conclude that the spin-orbit coupling $\Delta_R$ expected from symmetry arguments\cite{KM05}, has an curvature-intrinsic part besides the expected Rashba coupling due to an electric field $\Delta_R=\Delta_{\cal E}+\Delta_{\rm curv}$. Therefore $\Delta_R$ can be higher for curved graphene than for flat graphene. To our knowledge, this is the first time that this type of ``new'' spin-orbit coupling has been noticed. $\Delta_{\rm curv}$ is in a sense a new ``intrinsic/topological'' type of spin-orbit interaction in graphene which would be present even if $\cal E=$ 0,  as long as the samples present some type of corrugation. One important question now is how these ripples could affect macroscopic quantities. It has been already suggested that these ripples may be responsible for the lack of weak (anti)localization in graphene \cite{Metal06,MG06}. Other interesting macroscopic  quantities involving not only the pseudo spin but also the electronic spin maybe worth investigation. These issues are beyond the scope of the present paper and will be subject of future work.
   
It is also noteworthy that our estimates $\Delta_R \gg \Hi$, is opposite to the condition $\Delta_R \ll \Hi$ obtained by Kane and Mele\cite{KM05} to achieve the quantum spin Hall effect in graphene. So the quantum spin Hall effect may be achieved in neutral graphene ($\cal E=$ 0)  only below $\sim 0.01$K and provided the sample is also free of ripples so the curvature spin-orbit coupling $\Delta_{\rm curv} \ll \Hi$. Further progress in sample preparation seems needed to achieve such conditions although some preliminary improvements have been recently reported\cite{Metal06}. Moreover, corrugations in graphene could be seen as topological disorder. It has been shown that the spin Hall effect survives even if the spin-orbit gap $\Hi$ is closed by disorder\cite{SHMSM06}. A detailed discussion of the effect of disorder on the other two spin-orbit couplings $\Delta_{\cal E},\Delta_{\rm curv}$ will be presented elsewhere.  

The continuum model derived from microscopic parameters has been applied to
situations where also long range curvature effects can be significant. We have made estimates of the
effects of the various spin orbit terms on the low energy states of
fullerenes, nanotubes, and nanotube caps. For both nanotubes and nanotube caps we find that $\Delta_R \sim 1$K. For nanotubes we have clarified the existent discussion and reproduced the known appearance of a gap for the $n=0$ states and spin-splitting for $n \neq 0$ states in the energy spectrum. For nanotube caps states, we obtain indications that spin-orbit coupling may lead to spin-dependent emission possibilities for field-effect emission devices. This aspect will be investigated in the future.

\emph{Note added}. At the final stages of the writting of the present paper,
two preprints \cite{YAOYEFANG06,HMING06} have appeared. In these papers
similar estimates for $\Hi \sim 10^{-3}$meV  have been obtained for the
intrinsic spin-orbit coupling. Moreover similar discussion for the effect of
an perpendicular electric field has been also discussed in\cite{HMING06}. 
Our approach is similar to that followed in\cite{HMING06}. The two studies overlap and  
although the model used for the $\sigma$ band differs somewhat, the results are quantitative in agreement 
$\Delta_{\cal E} \sim 10^{-2}$meV

The two preprints\cite{YAOYEFANG06,HMING06} and our work agree in the estimation of
the intrinsic coupling, which turns out to be weak at the range of
temperatures of experimental interest (note, however, that we do not consider here
possible renormalization effects of this contribution\cite{KM05,GGV94b}).

On the other hand, the effect of local curvature $\Delta_{\rm curv}$ on the spin-orbit coupling has not been investigated in \cite{YAOYEFANG06,HMING06}. We show here that this term $\Delta_{\rm curv}$ is as important as, or perhaps even more important than the spin-orbit coupling due to an electric field $\Delta_{\cal E}$ for the typical values of $\cal E$ reported.

\section{Acknowledgements.}
We thank Yu. V. Nazarov, K. Novoselov, L. Brey, G. G\'omez Santos, Alberto Cortijo and Maria A. H. Vozmediano for valuable discussions. D.H-H and A.B. acknowledge funding from the Research Council of Norway, through grants no 162742/v00, 1585181/431 and 1158547/431.  F. G. acknowledges
funding from MEC (Spain) through grant FIS2005-05478-C02-01 and the European
Union Contract 12881 (NEST).

\appendix

\section{Appendix A. Two parameter analytical fit to the sigma bands of graphene.}
A simple approximation to the sigma bands of graphene takes only into account
the positions of the 2s and 2p atomic levels, $\epsilon_s$ and $\epsilon_p$,
and the interaction between nearest neighbor sp$^2$ orbitals. The three
sp$^2$ orbitals are:
\begin{eqnarray}
| 1 \rangle &\equiv &\frac{1}{\sqrt{3}} \left( | s \rangle + \sqrt{2} | p_x
  \rangle \right) \nonumber \\
| 2 \rangle &\equiv &\frac{1}{\sqrt{3}} \left[ | s \rangle + \sqrt{2} \left(
  - \frac{1}{2} | p_x \rangle + \frac{\sqrt{3}}{2} | p_y \rangle \right)
  \right] \nonumber \\
| 3 \rangle &\equiv &\frac{1}{\sqrt{3}} \left[ | s \rangle + \sqrt{2} \left(
  - \frac{1}{2} | p_x \rangle - \frac{\sqrt{3}}{2} | p_y \rangle \right)
  \right]
\end{eqnarray}
The two hopping elements considered are:
\begin{eqnarray}
V_1 &= &\langle i | {\cal H}_{\rm atom} | j \rangle |_{i \ne j} =
\frac{\epsilon_s - \epsilon_p}{3} \nonumber \\
V_2 &= &\langle i , m | {\cal H}_{\rm hopping} | i , n \rangle |_{ m , n: nearest-neighbors} = \frac{V_{ss \sigma} + 2 \sqrt{2} V_{sp \sigma} + 2 V_{pp
    \sigma}}{3}
\end{eqnarray}
where $i,j = 1,2,3$ denote ``bonding'' sp$^2$ states and $n,m$ denote atomic sites. $V_1$ depends on the geometry/angle between the bonds at each atom and $V_2$ depends on the coordination of nearest neighbors in the lattice. $V_1$ and $V_2$ therefore determine the details of band structure for the $\sigma$ bands\cite{TW71}. The energy associated with each ``bonding'' state $\langle i | {\cal H}_{\rm atom} | i \rangle|_{i =1,2,3}  = \left( \epsilon_s +2  \epsilon_p \right)/3$ is an energy constant independent of these details and not important for our discussion here.
 
 We label $a_1 , a_2 , a_3$ the amplitudes of a Bloch state on the three
orbitals at a given atom, and $\{ b_1 , b_2 , b_3 \} , \{ b'_1 , b'_2 , b'_3
\} , \{ b''_1 , b''_2 , b''_3 \}$ the amplitudes at its three nearest
neighbors. These amplitudes satisfy:
\begin{eqnarray}
\epsilon a_1 &= &V_1 ( a_2 + a_3 ) + V_2 b_1 \nonumber \\
\epsilon a_2  &= &V_1 ( a_1 + a_3 ) + V_2 b'_2 \nonumber \\
\epsilon a_3  &= &V_1 ( a_1 + a_2 ) + V_2 b''_3 \nonumber \\
\epsilon b_1 &= &V_1 ( b_2 + b_3 ) + V_2 a_1 \nonumber \\
\epsilon b'_2 &= &V_1 ( b'_1 + b'_3 ) + V_2 a_2 \nonumber \\
\epsilon b''_3 &= &V_1 ( b''_1 + b''_2 ) + V_2 a_3 
\label{amplitudes_sigma}
\end{eqnarray}
We can define two numbers, $a_n = a_1 + a_2 + a_3$ and $b_n = b_1 + b'_2 + b''_3$
associated to atom $n$. From Eq.(\ref{amplitudes_sigma}) we obtain:
\begin{eqnarray}
( \epsilon - 2 V_1 ) a_n &= &V_2 b_n \nonumber \\
( \epsilon + V_1 ) b_n &= &V_2 a_n + V_1 \sum_{n' ; n.-n.} a_{n'},
\end{eqnarray}
where $\sum_{n'; n.-n.} a_{n'}=(b_1+b_2+b_3)+(b'_1+b'_2+b'_3)+(b''_1+b''_2+b''_3)$ and $n.-n.$ denotes nearest-neighbors.

This equation is equivalent to:
\begin{equation}
\left( \epsilon - 2 V_1 - \frac{V_2^2}{\epsilon + V_1} \right) a_n =
\frac{V_1 V_2}{\epsilon + V_1} \sum_{n' ; n.-n.} a_{n'}
\label{amplitudes_eigenstates}
\end{equation}
Hence, the amplitudes $a_n$ satisfy an equation formally identical to the
tight binding equations for a single orbital model with nearest neighbor
hoppings in the honeycomb lattice. In momentum space, we can write:
\begin{equation}
\left( \epk - 2 V_1 - \frac{V_2^2}{\epk + V_1} \right) = \pm \frac{V_1 V_2}{\epk +
V_1} f_{{\vk}}
\label{amplitudes_sigma_2}
\end{equation}
where:
\begin{equation}
f_{\vk} \equiv \sqrt{ 3 + 2 \cos ( {\vk}  \vec { \bf a}_1 ) + 2 \cos ( {\vk} {\bf
  \vec{a}_2} ) + 2 \cos [ {\vk} (   \vec { \bf a}_1 -  \vec {\bf a}_2 )]}
\end{equation}
and $ \vec { \bf a}_1 , \vec {\bf a}_2$ are the unit vectors of the
honeycomb lattice. 

The derivation of equation(\ref{amplitudes_sigma_2})
 assumes that $a_n \ne 0$. There are
also solutions to eqs.(\ref{amplitudes_sigma}) for which $a_n = 0$ at all
sites. These solutions, and Eq.(\ref{amplitudes_sigma_2}) lead to:
\begin{eqnarray}
\epk &= &\frac{V_1}{2} \pm \sqrt{\frac{9 V_1^2}{4} + V_2^2 \pm V_1 V_2 f_{{\vk}}}
\nonumber \\
\epk &= &-V_1 \pm V_2
\label{sigma_bands}
\end{eqnarray}
These equations give the six $\sigma$ bands used in the main text.

In order to calculate the effects of transitions between the $\pi$ band and
the $\sigma$ band on the spin orbit coupling, we also need the matrix
elements of the spin orbit interaction at the points $K$ and $K'$. At the $K$
point, for instance, the hamiltonian for the $\sigma$ band is:
\begin{equation}
{\cal H}_{\sigma K} \equiv \left( \begin{array}{cccccc} 0 &V_1 &V_1 &V_2 &0 &0
  \\ V_1 &0 &V_1 &0 &V_2 e^{ 2 \pi i / 3} &0 \\ V_1 &V_1 &0 &0 &0 &V_2 e^{4
  \pi i / 3} \\ V_2 &0 &0 &0 &V_1 &V_1 \\ 0 &V_2 e^{4 \pi i / 3} &0 &V_1 &0
  &V_1 \\ 0 &0 &V_2 e^{2 \pi i / 3} &V_1 &V_1 &0 \end{array} \right)
\label{hamil_sigma} 
\end{equation}
The knowledge of the eigenstates, Eq.(\ref{amplitudes_eigenstates}) allows us to obtain
also the eigenvalues of Eq.(\ref{hamil_sigma}). The spin orbit coupling
induces transitions from the $| K , A , \uparrow \rangle$ state to the sigma
bands with energies $V_1 \pm V_2$ and spin down, and from the  $| K , A , \downarrow
\rangle$ state to the sigma bands with energies $V_1 / 2 \pm \sqrt{(9
  V_1^2)/4 + V_2^2}$ and spin up. The inverse processes are induced for Bloch
states localized at sublattice $B$. 

In the limit $V_1 \ll V_2$, the $\sigma$ bands lie at energies $\pm V_2$,
with corrections associated to $V_1$. The spin orbit coupling induces
transitions to the upper and lower bands, which tend to cancel. In addition,
the net effective intrinsic spin orbit coupling is the difference between the corrections to the
up spin bands minus the those for the down spin bands. The final effect is
that the strength of the intrinsic spin orbit coupling scales as $( \Delta^2
/ V_1 ) ( V_1 / V_2 )^4$ in the limit $V_1 \ll V_2$.

\section{Appendix B. Matrix elements of the spin orbit interaction in a sphere.}
Both the coupling induced by the curvature, 
eqs.(\ref{Rashba_K},\ref{Rashba_K'}), and the intrinsic coupling,
Eq.(\ref{intrinsic}), can be written, in a simple form, in a local basis of
wavefunctions where the spin is oriented perpendicular to the graphene
sheet, $| \theta , \phi , \perp \uparrow \rangle , | \theta , \phi , \perp \downarrow \rangle$. 
Using spherical coordinates, $\theta$ and $\phi$, the basis where the spins
are oriented parallel to the $z$ axis, $| \uparrow \rangle , | \downarrow
\rangle$,
can be written as:
\begin{eqnarray}
| \uparrow \rangle &\equiv &\cos \left( \frac{\theta}{2} \right) e^{i \phi/2} |
\theta , \phi , \perp \uparrow \rangle - \sin  \left( \frac{\theta}{2}
\right) e^{+ i \phi/2} |
\theta , \phi , \perp \downarrow  \rangle \nonumber \\ | \downarrow \rangle &\equiv & \sin  \left(
  \frac{\theta}{2} \right) e^{- i \phi/2} |  \theta , \phi , \perp \uparrow \rangle + \cos \left(
  \frac{\theta}{2} \right)  e^{-i \phi/2} | \theta , \phi , \perp \downarrow \rangle
\label{local_spin}
\end{eqnarray}
where the states $| + \rangle$ and $| - \rangle$ are defined in terms of some
fixed frame of reference. From this expression, we find in the basis $\{ |A \uparrow \rangle ,|A \downarrow \rangle,|B \uparrow \rangle ,|B \downarrow \rangle,$ basis for ${\cal K}$:
\begin{equation}
{\cal H}^{{\cal K}}_{\rm S-O} = \left( \begin{array}{cccc} 
    \Hi  \cos ( \theta ) & \Hi  \sin ( \theta ) e^{-i\phi} &  \HRm \sin \left( \frac{\theta}{2} \right) \cos \left( \frac{\theta}{2} \right) & \HR \cos^2 \left( \frac{\theta}{2} \right) e^{-i \phi}\\ 
     \Hi  \sin ( \theta ) e^{i\phi} & -\Hi  \cos ( \theta )& \HRm \sin^2 \left( \frac{\theta}{2} \right) e^{+ i \phi} &\HR \sin \left( \frac{\theta}{2} \right) \cos 
 \left( \frac{\theta}{2} \right) \\ 
     \HR \sin \left( \frac{\theta}{2} \right) \cos \left( \frac{\theta}{2} \right)&  \HR \sin^2 \left( \frac{\theta}{2} \right) e^{- i \phi} &-\Hi  \cos ( \theta ) & -\Hi  \sin ( \theta ) e^{-i\phi} \\
     \HRm \cos^2 \left( \frac{\theta}{2} \right) e^{i \phi}& \HRm \sin \left( \frac{\theta}{2} \right) \cos 
 \left( \frac{\theta}{2} \right) &-\Hi  \sin ( \theta ) e^{i\phi} & \Hi  \cos ( \theta )\\
\end{array} \right)
\end{equation}
and for ${\cal K'}$
\begin{equation}
{\cal H}^{{\cal K'}}_{\rm S-O} = \left( \begin{array}{cccc}    -\Hi  \cos ( \theta ) & -\Hi  \sin ( \theta ) e^{-i\phi} &\HR \sin \left( \frac{\theta}{2} \right) \cos \left( \frac{\theta}{2} \right)&\HR \sin^2 \left( \frac{\theta}{2} \right) e^{-i \phi} \\
     -\Hi  \sin ( \theta ) e^{i\phi} & \Hi  \cos ( \theta )& \HRm \cos^2 \left( \frac{\theta}{2} \right) e^{+ i \phi} &\HRm \sin \left( \frac{\theta}{2} \right) \cos 
 \left( \frac{\theta}{2} \right)\\
      \HRm \sin \left( \frac{\theta}{2} \right) \cos \left( \frac{\theta}{2} \right)&  \HR \cos^2 \left( \frac{\theta}{2} \right) e^{- i \phi} &\Hi  \cos ( \theta ) & \Hi  \sin ( \theta ) e^{-i\phi} \\ 
      \HRm \sin^2 \left( \frac{\theta}{2} \right) e^{i \phi}& \HR \sin \left( \frac{\theta}{2} \right) \cos 
 \left( \frac{\theta}{2} \right) & \Hi  \sin ( \theta ) e^{i\phi} & -\Hi  \cos ( \theta ) \\ 
    \end{array} \right)
 \label{H_fullerenes_K_1}
\end{equation}

\section{Appendix C. Matrix elements of the spin orbit interaction in a cylinder.}
The previous continuum analysis can be extended to nanotubes. We use cylindrical
coordinates, $z , \phi$, and, as before, define the spin orientations $|
\uparrow \rangle , | \downarrow \rangle$ as parallel and antiparallel to the
$z$ axis. The matrix elements can be obtained in a similar way
to Eq.(\ref{H_fullerenes_K_1}) by choosing $\theta =\pi/2$.

\begin{equation}
{\cal H}^{{\cal K}}_{\rm S-O} = \left( \begin{array}{cccc} 
    0 & \Hi  e^{-i\phi} &  \HRm /2 & \HR /2 e^{-i \phi}\\ 
     \Hi  e^{i\phi} & 0& \HRm /2 e^{+ i \phi} &\HR /2 \\ 
     \HR /2&  \HR /2 e^{- i \phi} &0 & -\Hi  e^{-i\phi} \\
     \HRm /2 e^{i \phi}& \HRm/2 &-\Hi e^{i\phi} & 0\\
\end{array} \right)
\end{equation}
and for ${\cal K'}$
\begin{equation}
{\cal H}^{{\cal K'}}_{\rm S-O} = \left( \begin{array}{cccc}    0 & -\Hi  e^{-i\phi} &\HR /2&\HR /2 e^{-i \phi} \\
     -\Hi  e^{i\phi} & 0& \HRm /2 e^{+ i \phi} &\HRm/2 \\
      \HRm /2&  \HR /2 e^{- i \phi} &0 & \Hi  e^{-i\phi} \\ 
      \HRm /2 e^{i \phi}& \HR/2  & \Hi e^{i\phi} & 0 \\ 
    \end{array} \right)
 \label{H_fullerenes_K_2}
\end{equation}

After integrating over the nanotube circunference $\int d\phi$ the Hamiltonian above becomes:

\begin{equation}
{\cal H}^{{\cal K}}_{\rm S-O} = \left( \begin{array}{cccc} 
    0 & 0&  \HRm \pi & 0\\ 
     0& 0& 0 &\HR \pi \\ 
     \HR \pi&  0 &0 & 0 \\
     0& \HRm \pi &0 & 0\\
\end{array} \right)
\end{equation}
and for ${\cal K'}$
\begin{equation}
{\cal H}^{{\cal K'}}_{\rm S-O} = \left( \begin{array}{cccc}    
     0 & 0&\HR \pi &0 \\
     0 & 0& 0 &\HRm \pi \\
      \HRm \pi& 0 &0 & 0 \\ 
      0& \HR \pi  & 0 & 0 \\ 
    \end{array} \right)
 \label{H_fullerenes_K_33}
\end{equation}

\bibliography{SO_TB_CURV_PUBLISHED}

\newcommand{\npb}{Nucl. Phys.}\newcommand{\adv}{Adv.
  Phys.}\newcommand{\epl}{Europhys. Lett.}
\begin{thebibliography}{56}
\expandafter\ifx\csname natexlab\endcsname\relax\def\natexlab#1{#1}\fi
\expandafter\ifx\csname bibnamefont\endcsname\relax
  \def\bibnamefont#1{#1}\fi
\expandafter\ifx\csname bibfnamefont\endcsname\relax
  \def\bibfnamefont#1{#1}\fi
\expandafter\ifx\csname citenamefont\endcsname\relax
  \def\citenamefont#1{#1}\fi
\expandafter\ifx\csname url\endcsname\relax
  \def\url#1{\texttt{#1}}\fi
\expandafter\ifx\csname urlprefix\endcsname\relax\def\urlprefix{URL }\fi
\providecommand{\bibinfo}[2]{#2}
\providecommand{\eprint}[2][]{\url{#2}}

\bibitem[{\citenamefont{Wallace}(1947)}]{W47}
\bibinfo{author}{\bibfnamefont{P.~R.} \bibnamefont{Wallace}},
  \bibinfo{journal}{Phys. Rev.} \textbf{\bibinfo{volume}{71}},
  \bibinfo{pages}{622} (\bibinfo{year}{1947}).

\bibitem[{\citenamefont{Slonczewski and Weiss}(1958)}]{SW58}
\bibinfo{author}{\bibfnamefont{J.~C.} \bibnamefont{Slonczewski}}
  \bibnamefont{and} \bibinfo{author}{\bibfnamefont{P.~R.} \bibnamefont{Weiss}},
  \bibinfo{journal}{Phys. Rev.} \textbf{\bibinfo{volume}{109}},
  \bibinfo{pages}{272} (\bibinfo{year}{1958}).

\bibitem[{\citenamefont{DiVincenzo and Mele}(1984)}]{mele}
\bibinfo{author}{\bibfnamefont{D.~P.} \bibnamefont{DiVincenzo}}
  \bibnamefont{and} \bibinfo{author}{\bibfnamefont{E.~J.} \bibnamefont{Mele}},
  \bibinfo{journal}{Phys.Rev.B} \textbf{\bibinfo{volume}{29}},
  \bibinfo{pages}{1685} (\bibinfo{year}{1984}).

\bibitem[{\citenamefont{Novoselov et~al.}(2004)\citenamefont{Novoselov, Geim,
  Morozov, Jiang, Zhang, Dubonos, Gregorieva, and Firsov}}]{Netal04}
\bibinfo{author}{\bibfnamefont{K.~S.} \bibnamefont{Novoselov}},
  \bibinfo{author}{\bibfnamefont{A.~K.} \bibnamefont{Geim}},
  \bibinfo{author}{\bibfnamefont{S.~V.} \bibnamefont{Morozov}},
  \bibinfo{author}{\bibfnamefont{D.}~\bibnamefont{Jiang}},
  \bibinfo{author}{\bibfnamefont{Y.}~\bibnamefont{Zhang}},
  \bibinfo{author}{\bibfnamefont{S.~V.} \bibnamefont{Dubonos}},
  \bibinfo{author}{\bibfnamefont{I.~V.} \bibnamefont{Gregorieva}},
  \bibnamefont{and} \bibinfo{author}{\bibfnamefont{A.~A.}
  \bibnamefont{Firsov}}, \bibinfo{journal}{Science}
  \textbf{\bibinfo{volume}{306}}, \bibinfo{pages}{666} (\bibinfo{year}{2004}).

\bibitem[{\citenamefont{Novoselov
  et~al.}(2005{\natexlab{a}})\citenamefont{Novoselov, Geim, Morozov, Jiang,
  Katsnelson, Grigorieva, Dubonos, and Firsov}}]{Netal05}
\bibinfo{author}{\bibfnamefont{K.~S.} \bibnamefont{Novoselov}},
  \bibinfo{author}{\bibfnamefont{A.~K.} \bibnamefont{Geim}},
  \bibinfo{author}{\bibfnamefont{S.~V.} \bibnamefont{Morozov}},
  \bibinfo{author}{\bibfnamefont{D.}~\bibnamefont{Jiang}},
  \bibinfo{author}{\bibfnamefont{M.~I.} \bibnamefont{Katsnelson}},
  \bibinfo{author}{\bibfnamefont{I.~V.} \bibnamefont{Grigorieva}},
  \bibinfo{author}{\bibfnamefont{S.~V.} \bibnamefont{Dubonos}},
  \bibnamefont{and} \bibinfo{author}{\bibfnamefont{A.~A.}
  \bibnamefont{Firsov}}, \bibinfo{journal}{Nature}
  \textbf{\bibinfo{volume}{438}}, \bibinfo{pages}{197}
  (\bibinfo{year}{2005}{\natexlab{a}}).

\bibitem[{\citenamefont{Novoselov
  et~al.}(2005{\natexlab{b}})\citenamefont{Novoselov, Jiang, Schedin, Booth,
  Khotkevich, Morozov, and Geim}}]{Netal05b}
\bibinfo{author}{\bibfnamefont{K.~S.} \bibnamefont{Novoselov}},
  \bibinfo{author}{\bibfnamefont{D.}~\bibnamefont{Jiang}},
  \bibinfo{author}{\bibfnamefont{F.}~\bibnamefont{Schedin}},
  \bibinfo{author}{\bibfnamefont{T.~J.} \bibnamefont{Booth}},
  \bibinfo{author}{\bibfnamefont{V.~V.} \bibnamefont{Khotkevich}},
  \bibinfo{author}{\bibfnamefont{S.~V.} \bibnamefont{Morozov}},
  \bibnamefont{and} \bibinfo{author}{\bibfnamefont{A.~K.} \bibnamefont{Geim}},
  \bibinfo{journal}{Proc. Nat. Acad. Sc.} \textbf{\bibinfo{volume}{102}},
  \bibinfo{pages}{10451} (\bibinfo{year}{2005}{\natexlab{b}}).

\bibitem[{\citenamefont{Zhang et~al.}(2005)\citenamefont{Zhang, Tan, Stormer,
  and Kim}}]{Zetal05b}
\bibinfo{author}{\bibfnamefont{Y.}~\bibnamefont{Zhang}},
  \bibinfo{author}{\bibfnamefont{Y.-W.} \bibnamefont{Tan}},
  \bibinfo{author}{\bibfnamefont{H.~L.} \bibnamefont{Stormer}},
  \bibnamefont{and} \bibinfo{author}{\bibfnamefont{P.}~\bibnamefont{Kim}},
  \bibinfo{journal}{Nature} \textbf{\bibinfo{volume}{438}},
  \bibinfo{pages}{201} (\bibinfo{year}{2005}).

\bibitem[{\citenamefont{Neto et~al.}(2005)\citenamefont{Neto, Guinea, and
  Peres}}]{NGP05}
\bibinfo{author}{\bibfnamefont{A.~H.~C.} \bibnamefont{Neto}},
  \bibinfo{author}{\bibfnamefont{F.}~\bibnamefont{Guinea}}, \bibnamefont{and}
  \bibinfo{author}{\bibfnamefont{N.~M.~R.} \bibnamefont{Peres}}
  (\bibinfo{year}{2005}), \eprint{cond-mat/0509709}.

\bibitem[{\citenamefont{Peres et~al.}(2006)\citenamefont{Peres, Guinea, and
  {Castro Neto}}}]{PGN06}
\bibinfo{author}{\bibfnamefont{N.~M.~R.} \bibnamefont{Peres}},
  \bibinfo{author}{\bibfnamefont{F.}~\bibnamefont{Guinea}}, \bibnamefont{and}
  \bibinfo{author}{\bibfnamefont{A.~H.} \bibnamefont{{Castro Neto}}},
  \bibinfo{journal}{Phys. Rev. B} \textbf{\bibinfo{volume}{73}},
  \bibinfo{pages}{125411} (\bibinfo{year}{2006}).

\bibitem[{\citenamefont{Gusynin and Sharapov}(2005)}]{GS05}
\bibinfo{author}{\bibfnamefont{V.~P.} \bibnamefont{Gusynin}} \bibnamefont{and}
  \bibinfo{author}{\bibfnamefont{S.~G.} \bibnamefont{Sharapov}},
  \bibinfo{journal}{Phys. Rev. Lett.} \textbf{\bibinfo{volume}{95}},
  \bibinfo{pages}{146801} (\bibinfo{year}{2005}).

\bibitem[{\citenamefont{Brey and Fertig}(2006)}]{BF06}
\bibinfo{author}{\bibfnamefont{L.}~\bibnamefont{Brey}} \bibnamefont{and}
  \bibinfo{author}{\bibfnamefont{H.~A.} \bibnamefont{Fertig}},
  \bibinfo{journal}{Phys. Rev. B} \textbf{\bibinfo{volume}{73}},
  \bibinfo{pages}{195408} (\bibinfo{year}{2006}).

\bibitem[{\citenamefont{Khveshchenko}(2006{\natexlab{a}})}]{K05}
\bibinfo{author}{\bibfnamefont{D.~V.} \bibnamefont{Khveshchenko}},
  \bibinfo{journal}{Phys. Rev. Lett.} \textbf{\bibinfo{volume}{97}},
  \bibinfo{pages}{036802} (\bibinfo{year}{2006}{\natexlab{a}}).

\bibitem[{\citenamefont{Morpurgo and Guinea}(2006)}]{MG06}
\bibinfo{author}{\bibfnamefont{A.~F.} \bibnamefont{Morpurgo}} \bibnamefont{and}
  \bibinfo{author}{\bibfnamefont{F.}~\bibnamefont{Guinea}}
  (\bibinfo{year}{2006}), \eprint{cond-mat/0603789, Phys. Rev. Lett. (to be
  published)}.

\bibitem[{\citenamefont{Morozov et~al.}(2006)\citenamefont{Morozov, Novoselov,
  Katsnelson, Schedin, Jiang, and Geim}}]{Metal06}
\bibinfo{author}{\bibfnamefont{S.}~\bibnamefont{Morozov}},
  \bibinfo{author}{\bibfnamefont{K.}~\bibnamefont{Novoselov}},
  \bibinfo{author}{\bibfnamefont{M.}~\bibnamefont{Katsnelson}},
  \bibinfo{author}{\bibfnamefont{F.}~\bibnamefont{Schedin}},
  \bibinfo{author}{\bibfnamefont{D.}~\bibnamefont{Jiang}}, \bibnamefont{and}
  \bibinfo{author}{\bibfnamefont{A.~K.} \bibnamefont{Geim}},
  \bibinfo{journal}{Phys. Rev. Lett.} \textbf{\bibinfo{volume}{97}},
  \bibinfo{pages}{016801} (\bibinfo{year}{2006}).

\bibitem[{\citenamefont{McCann et~al.}(2006)\citenamefont{McCann, Kechedzhi,
  Fal'ko, Suzuura, Ando, and Altshuler}}]{MKFSAA06}
\bibinfo{author}{\bibfnamefont{E.}~\bibnamefont{McCann}},
  \bibinfo{author}{\bibfnamefont{K.}~\bibnamefont{Kechedzhi}},
  \bibinfo{author}{\bibfnamefont{V.~I.} \bibnamefont{Fal'ko}},
  \bibinfo{author}{\bibfnamefont{H.}~\bibnamefont{Suzuura}},
  \bibinfo{author}{\bibfnamefont{T.}~\bibnamefont{Ando}}, \bibnamefont{and}
  \bibinfo{author}{\bibfnamefont{B.~L.} \bibnamefont{Altshuler}},
  \bibinfo{journal}{Phys. Rev. Lett.} \textbf{\bibinfo{volume}{97}},
  \bibinfo{pages}{146805} (\bibinfo{year}{2006}).

\bibitem[{\citenamefont{Aleiner and Efetov}(2006)}]{AAEF06}
\bibinfo{author}{\bibfnamefont{I.~L.} \bibnamefont{Aleiner}} \bibnamefont{and}
  \bibinfo{author}{\bibfnamefont{K.~B.} \bibnamefont{Efetov}}
  (\bibinfo{year}{2006}), \eprint{cond-mat/0607200 (to be published)}.

\bibitem[{\citenamefont{Altland}(2006)}]{ALTH06}
\bibinfo{author}{\bibfnamefont{A.}~\bibnamefont{Altland}}
  (\bibinfo{year}{2006}), \eprint{cond-mat/0607247 (to be published)}.

\bibitem[{\citenamefont{Tworzydlo et~al.}(2006)\citenamefont{Tworzydlo,
  Trauzettel, Titov, Rycerz, and Beenakker}}]{TTTRB06}
\bibinfo{author}{\bibfnamefont{J.}~\bibnamefont{Tworzydlo}},
  \bibinfo{author}{\bibfnamefont{B.}~\bibnamefont{Trauzettel}},
  \bibinfo{author}{\bibfnamefont{M.}~\bibnamefont{Titov}},
  \bibinfo{author}{\bibfnamefont{A.}~\bibnamefont{Rycerz}}, \bibnamefont{and}
  \bibinfo{author}{\bibfnamefont{C.~W.~J.} \bibnamefont{Beenakker}},
  \bibinfo{journal}{Phys. Rev. Lett.} \textbf{\bibinfo{volume}{96}},
  \bibinfo{pages}{246802} (\bibinfo{year}{2006}).

\bibitem[{\citenamefont{Katsnelson et~al.}(2006)\citenamefont{Katsnelson,
  Novoselov, and Geim}}]{KNG06}
\bibinfo{author}{\bibfnamefont{M.~I.} \bibnamefont{Katsnelson}},
  \bibinfo{author}{\bibfnamefont{K.~S.} \bibnamefont{Novoselov}},
  \bibnamefont{and} \bibinfo{author}{\bibfnamefont{A.~K.} \bibnamefont{Geim}},
  \bibinfo{journal}{Nat. Phys.} \textbf{\bibinfo{volume}{2}},
  \bibinfo{pages}{620} (\bibinfo{year}{2006}).

\bibitem[{\citenamefont{Suzuura and Ando}(2002)}]{SA02}
\bibinfo{author}{\bibfnamefont{H.}~\bibnamefont{Suzuura}} \bibnamefont{and}
  \bibinfo{author}{\bibfnamefont{T.}~\bibnamefont{Ando}},
  \bibinfo{journal}{Phys. Rev. Lett.} \textbf{\bibinfo{volume}{89}},
  \bibinfo{pages}{266603} (\bibinfo{year}{2002}).

\bibitem[{\citenamefont{Nomura and MacDonald}(2006)}]{KNMcD06}
\bibinfo{author}{\bibfnamefont{K.}~\bibnamefont{Nomura}} \bibnamefont{and}
  \bibinfo{author}{\bibfnamefont{A.~H.} \bibnamefont{MacDonald}},
  \bibinfo{journal}{Phys. Rev. Lett.} \textbf{\bibinfo{volume}{96}},
  \bibinfo{pages}{256602} (\bibinfo{year}{2006}).

\bibitem[{\citenamefont{Khveshchenko}(2006{\natexlab{b}})}]{KhvE06}
\bibinfo{author}{\bibfnamefont{D.~V.} \bibnamefont{Khveshchenko}}
  (\bibinfo{year}{2006}{\natexlab{b}}), \eprint{cond-mat/0607174 (to be
  published)}.

\bibitem[{\citenamefont{Katsnelson}(2006)}]{Katsn06}
\bibinfo{author}{\bibfnamefont{M.~I.} \bibnamefont{Katsnelson}},
  \bibinfo{journal}{Eur. Phys. J. B} \textbf{\bibinfo{volume}{52}},
  \bibinfo{pages}{151} (\bibinfo{year}{2006}).

\bibitem[{\citenamefont{Kane and Mele}(2005)}]{KM05}
\bibinfo{author}{\bibfnamefont{C.~L.} \bibnamefont{Kane}} \bibnamefont{and}
  \bibinfo{author}{\bibfnamefont{E.~J.} \bibnamefont{Mele}},
  \bibinfo{journal}{Phys. Rev. Lett.} \textbf{\bibinfo{volume}{95}},
  \bibinfo{pages}{226801} (\bibinfo{year}{2005}).

\bibitem[{\citenamefont{Sheng et~al.}(2005)\citenamefont{Sheng, Sheng, Ting,
  and Haldane}}]{SSTH05}
\bibinfo{author}{\bibfnamefont{L.}~\bibnamefont{Sheng}},
  \bibinfo{author}{\bibfnamefont{D.~N.} \bibnamefont{Sheng}},
  \bibinfo{author}{\bibfnamefont{C.~S.} \bibnamefont{Ting}}, \bibnamefont{and}
  \bibinfo{author}{\bibfnamefont{F.~D.~M.} \bibnamefont{Haldane}},
  \bibinfo{journal}{Phys. Rev. Lett.} \textbf{\bibinfo{volume}{95}},
  \bibinfo{pages}{136602} (\bibinfo{year}{2005}).

\bibitem[{\citenamefont{Sinitsyn et~al.}(2006)\citenamefont{Sinitsyn, Hill,
  Min, Sinova, and MacDonald}}]{SHMSM06}
\bibinfo{author}{\bibfnamefont{N.~A.} \bibnamefont{Sinitsyn}},
  \bibinfo{author}{\bibfnamefont{J.~E.} \bibnamefont{Hill}},
  \bibinfo{author}{\bibfnamefont{H.}~\bibnamefont{Min}},
  \bibinfo{author}{\bibfnamefont{J.}~\bibnamefont{Sinova}}, \bibnamefont{and}
  \bibinfo{author}{\bibfnamefont{A.~H.} \bibnamefont{MacDonald}},
  \bibinfo{journal}{Phys. Rev. Lett.} \textbf{\bibinfo{volume}{97}},
  \bibinfo{pages}{106804} (\bibinfo{year}{2006}).

\bibitem[{\citenamefont{Esquinazi et~al.}(2003)\citenamefont{Esquinazi,
  Spemann, H\"ohne, Setzer, Han, and Butz}}]{esquinazi}
\bibinfo{author}{\bibfnamefont{P.}~\bibnamefont{Esquinazi}},
  \bibinfo{author}{\bibfnamefont{D.}~\bibnamefont{Spemann}},
  \bibinfo{author}{\bibfnamefont{R.}~\bibnamefont{H\"ohne}},
  \bibinfo{author}{\bibfnamefont{A.}~\bibnamefont{Setzer}},
  \bibinfo{author}{\bibfnamefont{K.-H.} \bibnamefont{Han}}, \bibnamefont{and}
  \bibinfo{author}{\bibfnamefont{T.}~\bibnamefont{Butz}},
  \bibinfo{journal}{Phys. Rev. Lett.} \textbf{\bibinfo{volume}{91}},
  \bibinfo{pages}{227201} (\bibinfo{year}{2003}).

\bibitem[{\citenamefont{Dresselhaus}(1955)}]{DSS55}
\bibinfo{author}{\bibfnamefont{G.}~\bibnamefont{Dresselhaus}},
  \bibinfo{journal}{Phys. Rev.} \textbf{\bibinfo{volume}{100}},
  \bibinfo{pages}{580} (\bibinfo{year}{1955}).

\bibitem[{\citenamefont{Slonczewski}(1955)}]{S55}
\bibinfo{author}{\bibfnamefont{J.~C.} \bibnamefont{Slonczewski}}, Ph.D. thesis,
  \bibinfo{school}{Rutgers University} (\bibinfo{year}{1955}).

\bibitem[{\citenamefont{Dresselhaus and Dresselhaus}(1965)}]{DD65}
\bibinfo{author}{\bibfnamefont{G.}~\bibnamefont{Dresselhaus}} \bibnamefont{and}
  \bibinfo{author}{\bibfnamefont{M.~S.} \bibnamefont{Dresselhaus}},
  \bibinfo{journal}{Phys. Rev.} \textbf{\bibinfo{volume}{A140}},
  \bibinfo{pages}{401} (\bibinfo{year}{1965}).

\bibitem[{\citenamefont{Bychkov and Rashba}(1984)}]{BYCHRASH84}
\bibinfo{author}{\bibfnamefont{Y.~A.} \bibnamefont{Bychkov}} \bibnamefont{and}
  \bibinfo{author}{\bibfnamefont{E.~I.} \bibnamefont{Rashba}},
  \bibinfo{journal}{J. Phys. C: Solid State Phys.}
  \textbf{\bibinfo{volume}{17}}, \bibinfo{pages}{6039} (\bibinfo{year}{1984}).

\bibitem[{\citenamefont{Y.Yao et~al.}(2006)\citenamefont{Y.Yao, Ye, Qi, Zhang,
  and Fang}}]{YAOYEFANG06}
\bibinfo{author}{\bibnamefont{Y.Yao}},
  \bibinfo{author}{\bibfnamefont{F.}~\bibnamefont{Ye}},
  \bibinfo{author}{\bibfnamefont{X.-L.} \bibnamefont{Qi}},
  \bibinfo{author}{\bibfnamefont{S.-C.} \bibnamefont{Zhang}}, \bibnamefont{and}
  \bibinfo{author}{\bibfnamefont{Z.}~\bibnamefont{Fang}}
  (\bibinfo{year}{2006}), \eprint{cond-mat/0606350 (to be published)}.

\bibitem[{\citenamefont{H.Min et~al.}(2006)\citenamefont{H.Min, Hill, Sinitsyn,
  Sahu, Kleinman, and MacDonald}}]{HMING06}
\bibinfo{author}{\bibnamefont{H.Min}}, \bibinfo{author}{\bibfnamefont{J.~E.}
  \bibnamefont{Hill}},
  \bibinfo{author}{\bibfnamefont{N.}~\bibnamefont{Sinitsyn}},
  \bibinfo{author}{\bibfnamefont{B.}~\bibnamefont{Sahu}},
  \bibinfo{author}{\bibfnamefont{L.}~\bibnamefont{Kleinman}}, \bibnamefont{and}
  \bibinfo{author}{\bibfnamefont{A.}~\bibnamefont{MacDonald}},
  \bibinfo{journal}{Phys. Rev. B.} \textbf{\bibinfo{volume}{74}},
  \bibinfo{pages}{165310} (\bibinfo{year}{2006}).

\bibitem[{\citenamefont{Thorpe and Weaire}(1971)}]{TW71}
\bibinfo{author}{\bibfnamefont{M.~F.} \bibnamefont{Thorpe}} \bibnamefont{and}
  \bibinfo{author}{\bibfnamefont{D.}~\bibnamefont{Weaire}},
  \bibinfo{journal}{Phys. Rev. Lett.} \textbf{\bibinfo{volume}{27}},
  \bibinfo{pages}{1581} (\bibinfo{year}{1971}).

\bibitem[{\citenamefont{Guinea}(1981)}]{G81}
\bibinfo{author}{\bibfnamefont{F.}~\bibnamefont{Guinea}}, \bibinfo{journal}{J.
  Phys. C: Condens. Matt.} \textbf{\bibinfo{volume}{14}}, \bibinfo{pages}{3345}
  (\bibinfo{year}{1981}).

\bibitem[{\citenamefont{Braensden and Joachain}(1983)}]{BJ83}
\bibinfo{author}{\bibfnamefont{B.~H.} \bibnamefont{Braensden}}
  \bibnamefont{and} \bibinfo{author}{\bibfnamefont{C.~J.}
  \bibnamefont{Joachain}}, \emph{\bibinfo{title}{Physics of Atoms and
  Molecules}} (\bibinfo{publisher}{Addison Wesley}, \bibinfo{year}{1983}).

\bibitem[{\citenamefont{Ando}(2000)}]{A00}
\bibinfo{author}{\bibfnamefont{T.}~\bibnamefont{Ando}}, \bibinfo{journal}{J.
  Phys. Soc. Jpn} \textbf{\bibinfo{volume}{69}}, \bibinfo{pages}{1757}
  (\bibinfo{year}{2000}).

\bibitem[{\citenamefont{Serrano et~al.}(2000)\citenamefont{Serrano, Cardona,
  and Ruf}}]{SCR00}
\bibinfo{author}{\bibfnamefont{J.}~\bibnamefont{Serrano}},
  \bibinfo{author}{\bibfnamefont{M.}~\bibnamefont{Cardona}}, \bibnamefont{and}
  \bibinfo{author}{\bibfnamefont{J.}~\bibnamefont{Ruf}},
  \bibinfo{journal}{Solid St. Commun.} \textbf{\bibinfo{volume}{113}},
  \bibinfo{pages}{411} (\bibinfo{year}{2000}).

\bibitem[{\citenamefont{Hermann and Skillman}(1963)}]{HS63}
\bibinfo{author}{\bibfnamefont{F.}~\bibnamefont{Hermann}} \bibnamefont{and}
  \bibinfo{author}{\bibfnamefont{S.}~\bibnamefont{Skillman}},
  \emph{\bibinfo{title}{Atomic Structure Calculations}}
  (\bibinfo{publisher}{Prentice Hall, Englewood Cliffs, NJ},
  \bibinfo{year}{1963}).

\bibitem[{\citenamefont{Tom\'anek and Louie}(1988)}]{TL88}
\bibinfo{author}{\bibfnamefont{D.}~\bibnamefont{Tom\'anek}} \bibnamefont{and}
  \bibinfo{author}{\bibfnamefont{S.~G.} \bibnamefont{Louie}},
  \bibinfo{journal}{Phys. Rev. B} \textbf{\bibinfo{volume}{37}},
  \bibinfo{pages}{8327} (\bibinfo{year}{1988}).

\bibitem[{\citenamefont{Tom\'anek and Schl\"uter}(1991)}]{TS91}
\bibinfo{author}{\bibfnamefont{D.}~\bibnamefont{Tom\'anek}} \bibnamefont{and}
  \bibinfo{author}{\bibfnamefont{M.~A.} \bibnamefont{Schl\"uter}},
  \bibinfo{journal}{Phys. Rev. Lett.} \textbf{\bibinfo{volume}{67}},
  \bibinfo{pages}{2331} (\bibinfo{year}{1991}).

\bibitem[{\citenamefont{Brandt et~al.}(1988)\citenamefont{Brandt, Chudinov, and
  Ponomarev}}]{BCP88}
\bibinfo{author}{\bibfnamefont{N.~B.} \bibnamefont{Brandt}},
  \bibinfo{author}{\bibfnamefont{S.~M.} \bibnamefont{Chudinov}},
  \bibnamefont{and} \bibinfo{author}{\bibfnamefont{Y.~G.}
  \bibnamefont{Ponomarev}}, in \emph{\bibinfo{booktitle}{Modern Problems in
  Condensed Matter Sciences}}, edited by \bibinfo{editor}{\bibfnamefont{V.~M.}
  \bibnamefont{Agranovich}} \bibnamefont{and}
  \bibinfo{editor}{\bibfnamefont{A.~A.} \bibnamefont{Maradudin}}
  (\bibinfo{publisher}{North Holland (Amsterdam)}, \bibinfo{year}{1988}), vol.
  \bibinfo{volume}{20.1}.

\bibitem[{\citenamefont{McClure and Yafet}(1962)}]{MY62}
\bibinfo{author}{\bibfnamefont{J.~W.} \bibnamefont{McClure}} \bibnamefont{and}
  \bibinfo{author}{\bibfnamefont{Y.}~\bibnamefont{Yafet}}, in
  \emph{\bibinfo{booktitle}{5th Conference on Carbon}}
  (\bibinfo{publisher}{Pergamon, University Park, Pennsylvania},
  \bibinfo{year}{1962}).

\bibitem[{\citenamefont{Yafet}(1963)}]{Y63}
\bibinfo{author}{\bibfnamefont{Y.}~\bibnamefont{Yafet}}, in
  \emph{\bibinfo{booktitle}{Solid State Physics}}, edited by
  \bibinfo{editor}{\bibfnamefont{F.}~\bibnamefont{Seitz}} \bibnamefont{and}
  \bibinfo{editor}{\bibfnamefont{D.}~\bibnamefont{Turnbull}}
  (\bibinfo{publisher}{Academic Press, New York}, \bibinfo{year}{1963}),
  vol.~\bibinfo{volume}{14}, p.~\bibinfo{pages}{2}.

\bibitem[{\citenamefont{McClure}(1957)}]{M57}
\bibinfo{author}{\bibfnamefont{J.~W.} \bibnamefont{McClure}},
  \bibinfo{journal}{Phys. Rev.} \textbf{\bibinfo{volume}{108}},
  \bibinfo{pages}{612} (\bibinfo{year}{1957}).

\bibitem[{\citenamefont{Kroto et~al.}(1985)\citenamefont{Kroto, Heath, O'Brien,
  Curl, and Smalley}}]{KROTO85}
\bibinfo{author}{\bibfnamefont{H.~W.} \bibnamefont{Kroto}},
  \bibinfo{author}{\bibfnamefont{J.~R.} \bibnamefont{Heath}},
  \bibinfo{author}{\bibfnamefont{S.~C.} \bibnamefont{O'Brien}},
  \bibinfo{author}{\bibfnamefont{R.~F.} \bibnamefont{Curl}}, \bibnamefont{and}
  \bibinfo{author}{\bibfnamefont{R.~E.} \bibnamefont{Smalley}},
  \bibinfo{journal}{Nature} \textbf{\bibinfo{volume}{318}},
  \bibinfo{pages}{162} (\bibinfo{year}{1985}).

\bibitem[{\citenamefont{Gonz\'alez et~al.}(1992)\citenamefont{Gonz\'alez,
  Guinea, and Vozmediano}}]{GGV92}
\bibinfo{author}{\bibfnamefont{J.}~\bibnamefont{Gonz\'alez}},
  \bibinfo{author}{\bibfnamefont{F.}~\bibnamefont{Guinea}}, \bibnamefont{and}
  \bibinfo{author}{\bibfnamefont{M.~A.~H.} \bibnamefont{Vozmediano}},
  \bibinfo{journal}{\prl} \textbf{\bibinfo{volume}{69}}, \bibinfo{pages}{172}
  (\bibinfo{year}{1992}).

\bibitem[{\citenamefont{Gonz\'alez et~al.}(1993)\citenamefont{Gonz\'alez,
  Guinea, and Vozmediano}}]{GGV93b}
\bibinfo{author}{\bibfnamefont{J.}~\bibnamefont{Gonz\'alez}},
  \bibinfo{author}{\bibfnamefont{F.}~\bibnamefont{Guinea}}, \bibnamefont{and}
  \bibinfo{author}{\bibfnamefont{M.~A.~H.} \bibnamefont{Vozmediano}},
  \bibinfo{journal}{Nucl. Phys. B} \textbf{\bibinfo{volume}{406 [FS]}},
  \bibinfo{pages}{771} (\bibinfo{year}{1993}).

\bibitem[{\citenamefont{Guinea et~al.}(1993)\citenamefont{Guinea, Gonz\'alez,
  and Vozmediano}}]{GGV93c}
\bibinfo{author}{\bibfnamefont{F.}~\bibnamefont{Guinea}},
  \bibinfo{author}{\bibfnamefont{J.}~\bibnamefont{Gonz\'alez}},
  \bibnamefont{and} \bibinfo{author}{\bibfnamefont{M.~A.~H.}
  \bibnamefont{Vozmediano}}, \bibinfo{journal}{Phys. Rev. B}
  \textbf{\bibinfo{volume}{47}}, \bibinfo{pages}{16576} (\bibinfo{year}{1993}).

\bibitem[{\citenamefont{DeMartino et~al.}(2002)\citenamefont{DeMartino, Egger,
  Hallberg, and Balseiro}}]{Metal02}
\bibinfo{author}{\bibfnamefont{A.}~\bibnamefont{DeMartino}},
  \bibinfo{author}{\bibfnamefont{R.}~\bibnamefont{Egger}},
  \bibinfo{author}{\bibfnamefont{K.}~\bibnamefont{Hallberg}}, \bibnamefont{and}
  \bibinfo{author}{\bibfnamefont{C.~A.} \bibnamefont{Balseiro}},
  \bibinfo{journal}{Phys. Rev. Lett.} \textbf{\bibinfo{volume}{88}},
  \bibinfo{pages}{206402} (\bibinfo{year}{2002}).

\bibitem[{\citenamefont{Chico et~al.}(2004)\citenamefont{Chico, L\'opez-Sancho,
  and Mu{\~n}oz}}]{CLM04}
\bibinfo{author}{\bibfnamefont{L.}~\bibnamefont{Chico}},
  \bibinfo{author}{\bibfnamefont{M.~P.} \bibnamefont{L\'opez-Sancho}},
  \bibnamefont{and} \bibinfo{author}{\bibfnamefont{M.~C.}
  \bibnamefont{Mu{\~n}oz}}, \bibinfo{journal}{Phys. Rev. Lett.}
  \textbf{\bibinfo{volume}{93}}, \bibinfo{pages}{176402}
  (\bibinfo{year}{2004}).

\bibitem[{\citenamefont{Wakabayashi and Sigrist}(2000)}]{WS00}
\bibinfo{author}{\bibfnamefont{K.}~\bibnamefont{Wakabayashi}} \bibnamefont{and}
  \bibinfo{author}{\bibfnamefont{M.}~\bibnamefont{Sigrist}},
  \bibinfo{journal}{\prl} \textbf{\bibinfo{volume}{84}}, \bibinfo{pages}{3390}
  (\bibinfo{year}{2000}).

\bibitem[{\citenamefont{Wakabayashi}(2001)}]{W01}
\bibinfo{author}{\bibfnamefont{K.}~\bibnamefont{Wakabayashi}},
  \bibinfo{journal}{\prb} \textbf{\bibinfo{volume}{64}},
  \bibinfo{pages}{125428} (\bibinfo{year}{2001}).

\bibitem[{\citenamefont{Yaguchi and Ando}(2001)}]{YA01}
\bibinfo{author}{\bibfnamefont{T.}~\bibnamefont{Yaguchi}} \bibnamefont{and}
  \bibinfo{author}{\bibfnamefont{T.}~\bibnamefont{Ando}},
  \bibinfo{journal}{Journ. Phys. Soc. Jpn.} \textbf{\bibinfo{volume}{70}},
  \bibinfo{pages}{1327} (\bibinfo{year}{2001}).

\bibitem[{\citenamefont{Novikov and Levitov}(2006)}]{FECAPS05}
\bibinfo{author}{\bibfnamefont{D.}~\bibnamefont{Novikov}} \bibnamefont{and}
  \bibinfo{author}{\bibfnamefont{L.}~\bibnamefont{Levitov}},
  \bibinfo{journal}{Phys. Rev. Lett.} \textbf{\bibinfo{volume}{96}},
  \bibinfo{pages}{036402} (\bibinfo{year}{2006}).

\bibitem[{\citenamefont{Gonz\'alez et~al.}(1994)\citenamefont{Gonz\'alez,
  Guinea, and Vozmediano}}]{GGV94b}
\bibinfo{author}{\bibfnamefont{J.}~\bibnamefont{Gonz\'alez}},
  \bibinfo{author}{\bibfnamefont{F.}~\bibnamefont{Guinea}}, \bibnamefont{and}
  \bibinfo{author}{\bibfnamefont{M.~A.~H.} \bibnamefont{Vozmediano}},
  \bibinfo{journal}{Nucl. Phys. B} \textbf{\bibinfo{volume}{424 [FS]}},
  \bibinfo{pages}{595} (\bibinfo{year}{1994}).

\end{thebibliography}


\end{document}